\documentclass{scrartcl}

\usepackage[authoryear]{natbib}
\usepackage{amsmath, amssymb, amsthm}
\usepackage[a4paper]{geometry}
\usepackage{booktabs}
\usepackage[toc,page]{appendix}
\usepackage{graphicx}
\usepackage[ruled]{algorithm2e}

\SetKwInput{KwInput}{Input}
\SetKwInput{KwOutput}{Output}
\SetKwFor{ParFor}{for}{do in parallel}{end for}
\usepackage{url}
\usepackage{siunitx}
\numberwithin{equation}{section}

\usepackage{multirow}
\theoremstyle{plain}

\newtheorem{definition}{Definition}

\renewcommand{\v}[1]{\boldsymbol{#1}}
\newcommand{\blind}{0} 

\def\*#1{\mathbf{#1}}

\usepackage{subcaption}

\usepackage[english]{babel}

\newcommand{\im}{\mathrm{i}}

\title{Time-Varying Multi-Seasonal ARMA Models}
\if0\blind
{
\author{Ganna Fagerberg$^{a}$\thanks{Corresponding author: ganna.fagerberg@stat.su.se. 
\noindent $^a$Department of Statistics, Stockholm University. 
$^b$School of Business, University of New South Wales. $^c$Data Analytics Center for Resources and Environments (DARE).}, Mattias Villani$^{a}$ and Robert Kohn$^{b,c}$}
}\fi
\date{}

\begin{document}      
\maketitle
\begin{abstract}
We propose an ARMA model that allows for multiple seasonal periods and time varying parameters in both regular and seasonal components, building upon previous work for pure AR processes and the conditional likelihood. The model is parameterized to ensure stability and invertibility at each time point. The parameter evolution is governed by dynamic shrinkage processes, enabling extended periods of essentially constant parameters, gradual changes, and abrupt shifts. The model includes a stochastic volatility component to account for potentially heterogeneous noise, also modeled by a dynamic shrinkage process. A Gibbs sampler is developed using the exact likelihood, with separate updating steps for the latent errors and the unobserved pre-sample history of the process. The time-varying AR and MA parameters are sampled jointly using a fast posterior sampler based on the extended Kalman filter. The model and the efficiency of the Gibbs sampler are evaluated using simulated and real data. A case study on monthly air passenger data in the US during 1990-2024 reveals significant changes in seasonality during the Covid-19 pandemic.

\end{abstract}

\section{Introduction}\label{sec:intro}

Autoregressive integrated moving average (ARIMA)  models and the Box-Jenkins modeling methodology \citep{boxjen@1970} is a comprehensive framework for modeling time series data. ARIMA models are known for their natural parsimony, particularly when seasonality is modeled using their elegant multiplicative structure. This structure readily extends to multiple seasonal periods, an increasingly common feature in modern time series data. 

The Box-Jenkins method relies on the restrictive assumption of \textit{global} stationarity, meaning that the process is time invariant, possibly after differencing. However, due to inherent process dynamics or external factors like pandemics, wars, and natural disasters, time series can experience both gradual and abrupt shifts, when studied over long stretches of time \citep{lubik2015time, giudici2023bayesian}. The common recommendation to  difference the series to achieve stationarity can lead to over-differencing, resulting in misleading inferences \citep{granger1980introduction} and poor forecasts \citep{makridakis1997arma, smith1994forecasting}. 

Instead of assuming global stationarity, it has been common to use models that are \textit{locally} stationary in a neighborhood around each time point, see e.g. \citet{dahlhaus2000likelihood}. Other popular classes of models have a small set of stationary regimes, with abrupt transitions between them, for example Markov switching type of models \citep{hamilton1989new, billio1999bayesian, steel2010bayesian},  change-point models \citep{chib1998estimation, rosen2012adaptspec}, and dynamic mixture models \citep{gerlach2000efficient}. Time-varying autoregressive (TVAR) models are also widely used \citep{prado2010time, yang2016bayesian, lubik2015time, wood2011bayesian}, in which the parameters evolve over time following a stochastic process, often a random walk with Gaussian innovations. 

Time-varying autoregressive moving average (TV-ARMA) models have not been as widely used as TV-AR models, likely due to the added complexity introduced by the MA components \citep{chan2017efficient, triantafyllopoulos2007bayesian}, even in the constant parameter case. First, the likelihood function is non-linear because the MA parameters must be estimated jointly with the unknown errors. Second, the MA parameters in Gaussian models may not be uniquely identified without imposing non-linear restrictions (\cite{chib1994bayes,marriott1993bayesian, chan2016large}). These complexities are further amplified in ARMA models, particularly in the time-varying case.

Several attempts have been made to estimate TV-ARMA models using a Bayesian approach. \cite{triantafyllopoulos2007bayesian} develops a framework for time-varying MA parameters by modeling the evolution model for the squares of the parameter, using conjugate update for estimation and forecasting.  \cite{huang2013time} propose a TV-ARMA model, using a sequential Monte Carlo method to track the time-varying parameters of the process under an $\alpha$-stable distribution. \cite{chan2017efficient} extend the TV model to the multivariate VARMA framework. 

There is a lack of time-varying ARMA models that explicitly account for seasonality. Modeling seasonality jointly with other model parameters can improve model accuracy and provide deeper insights into both the seasonal and non-seasonal dynamics of the data. \cite{fagerberg2026time} argue that the absence of time-varying seasonality in pure AR models is probably due to the non-linearity resulting from the multiplicative seasonal structure,  precluding the use of the Kalman filter for inference. \cite{fagerberg2026time} develop a time-varying seasonal AR process where both regular and seasonal AR parameters evolve over time. The time evolution in all the parameters is modeled using dynamic shrinkage priors \citep{kowal2019dynamic}, to allow for abrupt changes as well as periods with essentially constant parameters. Our article extends the framework of  \cite{fagerberg2026time} to include regular and seasonal MA parameters, and uses the exact likelihood instead of the conditional likelihood used by \cite{fagerberg2026time}. Additionally, the stochastic volatility component of the model is also modeled using a dynamic shrinkage process prior. We show that it is possible to extend the modeling framework developed in  \cite{fagerberg2026time} by sampling the unknown errors and the unobserved initial observations in separate Gibbs updating steps.

In static ARMA models, it is common to parameterize the AR and MA coefficients to guarantee stability and invertibility \citep{marriott1993bayesian, chib1994bayes, barnett1997robust}. Imposing these restrictions in time-varying parameter ARMA models is less common. The primary reason for this is likely computational: these restrictions are recursive and non-linear \citep{Barndorff-Nielsen1973, Monahan1984note}, which precludes using Kalman filter-based sampling algorithms \citep{carter1994gibbs, fruhwirth1994data} for inference. Nevertheless, a process with time-varying AR parameters  may drift into non-stable regions during certain time periods, leading to undesirable explosive forecast paths. Similarly, a time-varying MA process may enter non-invertible regions at certain points in time, resulting in a temporary lack of identifiability and interpretability leading to large fluctuations in the parameter estimates and poor forecast accuracy \citep{box2015time, wei2019time}. 

Our article makes the following contributions. First, a time-varying seasonal ARMA process is developed where both the regular and the seasonal ARMA parameters follow the dynamic shrinkage process priors in \citet{kowal2019dynamic}. Second, the process is made locally stable and  invertible by explicitly restricting the parameters to the stable and invertible region at every time point. Third, we propose a fully Bayesian approach based on the exact likelihood to jointly estimate the time-varying seasonal and non-seasonal ARMA coefficients of the seasonal TV-ARMA model, extending the Gibbs sampling algorithm in \cite{fagerberg2026time} with two additional steps - sampling of the latent error parameters and unobserved pre-sample lags.  Finally, we use dynamic shrinkage priors in a stochastic volatility model to capture potentially heterogeneous noise process. 

We explore the properties of the proposed model in several simulation experiments and compare it to the AdaptSpec method, which is often used as a benchmark model \citep{rosen2012adaptspec, BertolucciBayesSpec2021}. The model is applied to a time series of the monthly number of passengers between January 1990 and July 2024 for flights between New York and Miami. We find strong evidence of time-varying parameters, particularly in the seasonal MA parameters, where the otherwise strong seasonality is essentially wiped out during the Covid-19 pandemic.

Additional results to the ones in the paper are in the online supplementary material to this article, referenced below with the prefix S, e.g. Figure \ref{fig:usair_sma12_coef_stab} in Section \ref{app:usair_stability}.

\section{Time-varying ARMA models with multiple seasonal periods}

Sections \ref{subsec:TV-SARMA} and \ref{subsec:stability} introduce the time-varying multi-seasonal ARMA model with stability and invertibility constraints. The time-varying parameters are regularized using the dynamic shrinkage process (DSP) prior \citep{kowal2019dynamic}, as explained in Section \ref{subsec:sarma_dsp}. 

\subsection{Time-varying multi-seasonal ARMA model}\label{subsec:TV-SARMA}
The time-varying ARMA model with multiple seasonal components can be expressed as
\begin{equation}\label{eq: dynamic_ARMA}
   \phi_t(L) \prod_{j=1}^M\Phi_{jt}(L^{s_{j}})y_t =\psi_{t}(L)\prod_{j=1}^M\Psi_{jt}(L^{s_j})\varepsilon_t, 
\  \quad \varepsilon_t \overset{\mathrm{iid}}{\sim} N(0,\sigma_t^2)
\end{equation}
where \( L \) is the lag operator defined as $ L^k y_t = y_{t-k}$.  The factors $\phi_{t}(L) = 1-\phi_{1t} L^{1} -\phi_{2t} L^{2} -\ldots - \phi_{pt} L^{p}$ and $\Phi_{jt}(L^{s_j}) = 1-\Phi_{j1t} L^{s_j} -\Phi_{j2t} L^{2s_j} -\ldots - \Phi_{jP_jt} L^{P_js_j}$ denote the time-varying regular and seasonal AR polynomials, respectively, while $\psi_{t}(L) = 1+\psi_{1t} L^{1} +\psi_{2t} L^{2} +\ldots + \psi_{qt} L^{q}$ and $\Psi_{jt}(L^{s_j}) = 1+\Psi_{j1t} L^{s_j} +\Psi_{j2t} L^{2s_j} +\ldots + \Psi_{jQ_jt} L^{Q_js_j}$ denote the corresponding MA polynomials. Note that, for a given seasonal period, one of the AR or MA polynomials may have zero lags and can be removed from the model. The mean of the process is assumed to be zero throughout the paper, but can be modeled as a constant or a time-varying parameter.

We denote the model in \eqref{eq: dynamic_ARMA} by TV-SARMA$(p,\v P, q, \v Q)_{\v s}$, where $\v s = (s_1,\ldots, s_M)$ is the vector of $M$ seasonal periods; $p$ and $q$ are the nonseasonal AR and MA lag orders and $\v P = (P_1,\ldots,P_M)$ and $\v Q = (Q_1,\ldots,Q_M)$ are the corresponding AR and MA seasonal lag orders. 

Model \eqref{eq: dynamic_ARMA} is very flexible and can accommodate  purely non-seasonal as well as combinations of non-seasonal and seasonal processes with multiple seasonal periods in both AR and MA operators.  For example, when $M=0$, the model reduces to a non-seasonal ARMA($p,q$) model. When $M=1$, and  $\v s=s$, we obtain a single-seasonal ARMA specification 
\begin{equation}\label{eq: stat_AR_traditional}
    \phi_t(L) \Phi_t(L^s)y_t =  \psi_t(L) \Psi_t(L^s)\varepsilon_t.
\end{equation}
When both $p = 0$ and $\v P=0$, model  \eqref{eq: dynamic_ARMA} reduces to a pure time-varying multi-seasonal MA$(q, \v{Q})_{\v s}$ model. Conversely, when $q = 0$ and $\v Q=0$, the model simplifies to the pure time-varying multi-seasonal AR$(p, \v{P})_{\v s}$ process in \citet{fagerberg2026time}. 

\subsection{Imposing local stability and invertibility}\label{subsec:stability}

We impose stability conditions on all AR polynomials and invertibility conditions on all MA polynomials in  \eqref{eq: dynamic_ARMA}  at each time period $t$ of the process, using the parameterization in \citet{Barndorff-Nielsen1973} and \citet{Monahan1984note}. This maps a set of unconstrained coefficients to the set of AR and MA parameters that ensure stability and invertibility of the process. Since exactly the same map is used to enforce stability and invertibility (modulo a sign change), we first describe the stability restriction on the regular AR polynomial. Let \(\mathbb{S}^p \subset \mathbb{R}^p\) denote the region in parameter space where the \(\mathrm{AR}(p)\) process is stable. The reparameterization is defined as follows.

\begin{definition}[Stability]\label{def:stability_invertibility}
An \(\operatorname{ARMA}(p,q)\) process can be restricted to be stable through 1:1 and onto composite map from unrestricted AR parameters \(\boldsymbol{\theta} = (\theta_1, \ldots, \theta_p)^\top \in \mathbb{R}^p\) via the partial autocorrelations \(\boldsymbol{r} = (r_1, \ldots, r_p)^\top \in (-1, 1)^p\) to the stable AR parameters \(\boldsymbol{\phi} = (\phi_1, \ldots, \phi_p)^\top \in \mathbb{S}^p\). The mapping is defined as
\begin{equation*}
\v\theta \rightarrow \v r \rightarrow \boldsymbol{\phi},
\end{equation*}
where \(\phi_{1,1} = r_1\), and the recursion for \(k = 2, \ldots, p\) and \(j = 1, \ldots, k-1\) is
\begin{equation}\label{eq:stability_recursion}
\phi_{k,j} = \phi_{k-1,j} - r_k \phi_{k-1,k-j} 
\end{equation}
returning \(\v \phi = (\phi_{p,1}, \ldots, \phi_{p,p})^\top\), with \(\phi_{p,p} = r_p\).
\end{definition}
\noindent Exactly the same mapping can be used to restrict the ARMA process to be invertible; the only change is that the negative sign in \eqref{eq:stability_recursion} is replaced by a plus sign. 

We follow \cite{fagerberg2026time} and use the transformation suggested by \citet{Monahan1984note} to map the unrestricted parameters to the partial correlations, $\v\theta \rightarrow \v r$:
\begin{equation}\label{eq:t_transf}
 r_k=\frac{\theta_{k}}{\sqrt{1+\theta_{k}^2}}\quad  \text{ for } k=1,\ldots,m.
\end{equation}

In summary, to ensure both stationarity and invertibility at each  $t$, we map the unrestricted time-varying AR parameters $\v\theta_t = (\theta_{1t},\ldots, \theta_{pt})^\top$ to the time-varying stable AR coefficients  $\v\phi_{t}=(\phi_{1t},\ldots,\phi_{pt})^\top$, and the unrestricted time-varying MA parameters $\v\theta_t = (\theta_{1t},\ldots, \theta_{qt})^\top$ to the time-varying invertible MA coefficients $\v\psi_{t}=(\psi_{1t},\ldots,\psi_{qt})^\top$. This restriction is applied to each AR and MA polynomial in \eqref{eq: dynamic_ARMA}, but we suppress the polynomial index for notational clarity. 

The unrestricted parameters $\v\theta_t$ have no real interpretation, and we follow \citet{fagerberg2026time} and use a prior on $\v\theta_t$ that implies a uniform distribution on the AR parameters $\v\phi_t$ over the stability region. The same prior construction is used on the MA parameters to imply uniformity over the invertibility region, and on each seasonal AR and MA polynomial. 

A challenge in fitting ARMA models is the phenomenon of (near) root cancellation, which occurs when the polynomials $\phi(L)$ and $\psi(L)$ share one or more common roots, or when the roots are sufficiently close to each other \citep{box2015time}. If root cancellation occurs, some AR and MA parameters become redundant, resulting in non-identification in the likelihood. This can lead to  instability in the estimated coefficients, poor forecast performance, and imprecise impulse response functions \citep{chan2016large, kleibergen2000bayesian}. In TV-ARMA models, (near) root cancellation can occur at any time $t$. Following most of the ARMA literature, we assume that no root cancellation occurs, but Section \ref{subsec:experiment1} illustrates the effects of (near) root cancellation. 

\subsection{Dynamic shrinkage process for the AR and MA parameter evolutions}\label{subsec:sarma_dsp}
To complete the TV-SARMA model in \eqref{eq: dynamic_ARMA}, we let the time evolution of the unrestricted parameters $\v \theta_t$, as in \cite{fagerberg2026time}, have a dynamic shrinkage process (DSP) prior,  a class of global-local shrinkage priors introduced by \citet{kowal2019dynamic}. Our inference methodology is, however, agnostic to the specific choice of global-local shrinkage prior and can accommodate alternatives, for example, the dynamic triple Gamma prior in \citet{knaus2023dynamic}. The DSP is a time series extension of the widely used horseshoe prior \citep{carvalho2010horseshoe}, allowing the parameters to be essentially unchanged for long periods followed by large jumps or potentially persistent periods of rapid change. The process for the evolving TV-SARMA parameters for any given seasonality is

\begin{align}
\label{eq: evol_process}
   (\v\phi_t, \v\psi_t) &= \v g(\v\theta_{t}),  \nonumber \\
    \theta_{kt} &= \theta_{k,t-1} +\nu_{kt}, \hspace{3.45cm} 
    \nu_{kt}  \overset{\mathrm{ind}}{\sim} N\big(0, \exp(h_{kt})\big), \nonumber \\ 
    h_{kt} &= \mu_k + \kappa_k (h_{k,t-1}-\mu_k)+\eta_{kt}, \hspace{1cm}  
    {\eta}_{kt} \overset{\mathrm{iid}}{\sim} {Z}(1/2, 1/2, 0, 1), 
\end{align}
where $k=1,\ldots,p+q$ and \(Z\) denotes the heavy-tailed \(Z\)-distribution with zero location and unit scale parameters; see \citet{barndorff1982normal} and \citet{kowal2019dynamic} for details.  The transformation $(\v\phi_t, \v\psi_t) = \v g(\v\theta_{t})$ maps the $p+q$ unrestricted parameters $\v\theta_{t}=(\theta_{1t}, \ldots, \theta_{t,p+q})^\top \in \mathbb{R}^{p+q}$ to the stable AR parameters $\v\phi_t \in \mathbb{S}^p$ and the invertible MA parameters $\v\psi_t \in \mathbb{S}^q$ using the mapping in Definition \ref{def:stability_invertibility}; the mapping in Definition \ref{def:stability_invertibility} is applied to each seasonal polynomial separately.

 The overall degree of  time-variation in $\v \theta_t$ and, consequently, in $(\v\phi_t, \v\psi_t)$ is controlled by the global mean log variances $\v \mu = (\mu_1,\ldots,\mu_{p+q})^\top$. Local time-variation is driven by the local log variance innovations, $  {\eta}_{kt} \overset{\mathrm{iid}}{\sim} {Z}(1/2, 1/2, 0, 1)$, and the persistence of the log variance for the $k$th parameter, $h_{kt}$, is determined by $\kappa_k$. Section S2 in the supplementary material of \citet{fagerberg2026time} shows that the DSP properties on $\v\theta_t$   --- with parameters that can remain constant, have rapid movements or jumps --- carries over to the AR parameters $\v\phi_t$.

\section{Bayesian inference for time-varying seasonal ARMA}
The multi-seasonal TV-SARMA$(\mathbf{p}, \mathbf{P}, \mathbf{q}, \mathbf{Q})_\mathbf{s}$  in \eqref{eq: dynamic_ARMA} model can be written
\begin{equation}\label{eq: dynamic_ARMA_multiplied_out}
    y_t = \v x_{t}^\top \tilde{\v\phi}_{t} + \v z_{t}^\top \tilde{\v\psi}_{t}+\varepsilon_t,\quad \varepsilon_t \overset{\mathrm{iid}}{\sim} N(0,\sigma^2), 
\end{equation}
where the AR coefficients  $\tilde{\v\phi}_{t}$ and the MA coefficients $\tilde{\v\psi}_{t}$ contain all \emph {non-zero} coefficients  derived from multiplying out the AR and MA polynomials, $\tilde{\phi}_{t}(L) := \prod_{j=1}^M \phi_{jt}(L^{s_j})$ and $\tilde{\psi}_{t}(L):=\prod_{j=1}^M \psi_{jt}(L^{s_j})$, respectively. The vector $\v x^\top_{t}$  collects all lags $y_{t-k}$, $k=1,\ldots, p_{\max}$, for which the coefficients in $\tilde{\phi}_{t}(L)$ are non-zero. Similarly, $\v z_{t}$ collects all lags $ \varepsilon_{t-k}$, $k=1,\ldots, q_{\max}$, for which the coefficients in $\tilde{\psi}_{t}(L)$ are non-zero.   $p_{\max}$ and $q_{\max}$ are the maximum lag orders in $\v x_t^\top$ and $\v z_t^\top$,  respectively.  

As an example, consider a single-seasonal SARMA$(p=1,P=1,q=1,Q=1)_s$ model
\begin{equation}\label{eq: toy_ex}
    (1-\phi_{1t} L)(1-\Phi_{1t} L^s)y_t =  (1+\psi_{1t} L)(1+\Phi_{1t} L^s)\varepsilon_t.
\end{equation}
In the expanded form of \eqref{eq: dynamic_ARMA_multiplied_out}, the model can be written as:
\begin{equation}\label{eq: toy_additive}
    y_t = \phi_{1t} y_{t-1}+\Phi_{1t} y_{t-s} -\phi_{1t} \Phi_{1t} y_{t-(1+s)} +\psi_{1t} \varepsilon_{t-1}+\Psi_{1t} \varepsilon_{t-s} +\psi_{1t} \Psi_{1t} \varepsilon_{t-(1+s)} +\varepsilon_t, 
\end{equation}
where we collect the lagged observations in \( \v x_{t} = (y_{t-1}, y_{t-s}, y_{t-(1+s)})^\top \), the lagged errors in \( \v z_{t} = (\varepsilon_{t-1}, \varepsilon_{t-s}, \varepsilon_{t-(1+s)})^\top \), and the parameters in \( \tilde{\v\phi}_t = (\phi_{1t},\Phi_{1t}, -\phi_{1t}\Phi_{1t})^\top \) and \( \tilde{\v\psi}_t = (\psi_{1t},\Psi_{1t}, \psi_{1t}\Psi_{1t})^\top \).  Note that the parameter vector \( \tilde{\v\phi}_t \) for the AR part has three elements, but is a function of only two AR parameters, and the same is true for the MA part in \( \tilde{\v\psi}_t \). In general, a  TV-SARMA model with $M$ AR and MA polynomials can be viewed as a nonlinear transformation of totally $r = p+q+\sum_{j=1}^M (P_j +  Q_j)$ unrestricted time-varying AR and MA parameters.

The model in \eqref{eq: dynamic_ARMA_multiplied_out} is completed with the time evolution for the model parameters in \eqref{eq: evol_process}
\begin{align}
\label{eq: SARMA_DSP}
   y_t &= \v x_t^\top \tilde{\v\phi}_t + \v z_{t}^\top  \tilde{\v\psi}_t + \varepsilon_t, \hspace{2.9cm}  \varepsilon_t \overset{\mathrm{iid}}{\sim} N(0,\sigma^2) \nonumber \\
    ( \tilde{\v\phi}&_t , \tilde{\v\psi}_t) =  \tilde{\v g}(\v\theta_{t}) \nonumber  \\
    \v\theta_{t} &= \v\theta_{{t-1}} +\v\nu_{t}, \hspace{4.35cm} 
    \v\nu_{t}  \overset{\mathrm{ind}}{\sim} N\big(\mathbf{0}, \mathrm{Diag}(\exp(\v h_{t}))\big)  \nonumber \\ 
    {\v h}_{t} &= \v \mu + \v \kappa ({\v h}_{{t-1}}-\v \mu)+{\v\eta}_t, \hspace{2.25cm}   
    {\eta}_{kt} \overset{\mathrm{iid}}{\sim} {Z}(1/2,1/2, 0, 1), 
\end{align}
where $\v\theta_t$ is a vector with all $r$ AR and MA coefficients. The function $ ( \tilde{\v\phi}_t , \tilde{\v\psi}_t) = \v g(\v\theta_{t})$  maps the coefficients of each AR polynomial and MA polynomials, into the stability and invertibility regions, respectively, followed by the polynomial multiplication leading to  $\tilde{\v\phi}_{t}$ and $\tilde{\v\psi}_{t}$ in \eqref{eq: dynamic_ARMA_multiplied_out}.  The \( r \times r \) matrix \( \v\kappa = \mathrm{Diag}(\kappa_1, \ldots, \kappa_r)\) is diagonal, and \( \mathrm{Diag}(\exp(\mathbf{h}_t)) \) has \( \exp(\mathbf{h}_t) = (\exp(h_{1t}), \ldots, \exp(h_{rt}))^\top \)on its main diagonal.

\citet{fagerberg2026time} develop a Gibbs sampler for the special case of model \eqref{eq: SARMA_DSP} without MA terms, and use the conditional likelihood for inference by conditioning on pre-sample observations $\boldsymbol{x}_0=(y_0,y_{-1},\ldots,y_{-p_{\max}+1})^\top$. The full conditional posterior for the unrestricted AR parameters is sampled using the Forward Filtering and Backward Sampling (FFBS) method of \cite{carter1994gibbs} and \cite{fruhwirth1994data}, with the Kalman Filter replaced by the approximate Extended Kalman Filter (EKF); the algorithm is termed FFBSx with the x in FFBSx serving as a mnemonic for extended. \cite{fagerberg2026time} demonstrate that the FFBSx algorithm is fast, robust, and accurate for models like \eqref{eq: SARMA_DSP}; the accuracy of the sampler is verified by comparing the posterior from FFBSx to that from the simulation-consistent PGAS sampler \citep{lindsten2014particle}. \citet{fagerberg2026time}  argue that the reason for the accuracy of the sampler is that the non-linearity appears only in the observation equation and all innovations are (heteroscedastic) Gaussian conditional on the Polya-Gamma augmentation of the $Z$-distribution; see \citet{kowal2019dynamic}. The same properties hold for the posterior of the AR and MA parameters in the more general ARMA case here, since conditional on the error path $\varepsilon_{-q_{\max}:T}$ and the pre-sample observations \(\v x_0^\top\), the observation equation \eqref{eq: SARMA_DSP} is of the same non-linear regression type as in \citet{fagerberg2026time}.

We now extend the algorithm in \citet{fagerberg2026time} to the ARMA case with the exact likelihood function instead of the conditional likelihood. To preserve the general structure of the sampler, we add Gibbs updating steps to infer the entire trajectory $\varepsilon_{-q_{\max}:T}$ along with the vector of initial observations \(\v x_0^\top\) \citep{marriott1993bayesian, chib1994bayes, shaarawy1984bayesian}. This strategy allows treating these values as known inputs in \eqref{eq: SARMA_DSP}, thereby preserving the state-space structure of the model when updating $\v\theta_{0:T}$ and enabling direct use of the FFBSx algorithm in \cite{fagerberg2026time}. The Gibbs sampler is summarized in Algorithm~\ref{alg:gibbs_algorithm}. We will now detail the updating steps for the error paths $\varepsilon_{-q_{\max}:T}$,  the initial observations $\v x_0^\top$ and the noise process, $\sigma^2_{1:T}$ with DSP priors for the SV model.

\begin{algorithm} \label{alg: Gibbs_sampler}
\KwInput{data $y_{1:T}$}
\hspace{1.4cm}initial $\mu^{(0)}$, defaulting to $(\mu_0,\ldots,\mu_0)^\top$, where $\mu_0$ is the prior mean\\
\hspace{1.4cm}initial $\v h^{(0)}_{0:T}$, defaulting to $h_{kt}^{(0)}=\mu_0\text{ for all }k,t $\\
\hspace{1.4cm}initial $\sigma{^{2(0)}_{1:T}}$, defaulting to an estimate from the static seasonal ARMA \\
\hspace{1.4cm}initial $\v x_{0}{^{\top(0)}}$, defaulting to the mean of $y_t$, t=1,\ldots, 30  \\
\hspace{1.4cm}initial $\v z_{0}{^{\top(0)}}$, defaulting to $\varepsilon_{-({q_{maxl}-1}):0}=(0,\ldots,0)^\top$ , and $\varepsilon_{1:T}$ are computed recursively  \\
\hspace{1.4cm}the number of posterior draws $J$

\BlankLine
\For{$j = 1$ \KwTo $J$}{

    \BlankLine
    
    \textcolor{blue}{\textbackslash\textbackslash\hspace{0.1cm}draw the parameter evolutions using FFBSx} \\
    $\v \theta^{(j)}_{0:T}\leftarrow \texttt{FFBSx}(\v \theta_{0:T} \vert \v h^{(j-1)}_{0:T}, \sigma_{-(1:T}^{2(j-1)},{\varepsilon}^{(j-1)}_{-({q_{maxl}-1}):T}, \v x_{0}{^{\top(j-1)}},  y_{1:T})$ \\
    \BlankLine

\ParFor{$k = 1$ \KwTo $r$}{
\BlankLine

    \textcolor{blue}{\textbackslash\textbackslash\hspace{0.1cm}draw mixture of normals allocations} \\
   $\v a^{(j)}_{k,1:T} \leftarrow \ p(\v a_{k,1:T}\vert \v \theta^{(j)}_{k,0:T}, h^{(j-1)}_{k,0:T})$\\
   \BlankLine
   
    \textcolor{blue}{\textbackslash\textbackslash\hspace{0.1cm}draw log-volatilities} \\
   $\v h^{(j)}_{k,0:T} \leftarrow \ p(\v h_{k,0:T}\vert \v \theta^{(j)}_{k,0:T}, \v a^{(j)}_{k,1:T}, \kappa_k^{(j-1)}, \mu_k^{(j-1)}, \v\xi_{k, 1:T}^{(j-1)})$\\
   \BlankLine
   
    \textcolor{blue}{\textbackslash\textbackslash\hspace{0.1cm}draw P\'olya-Gamma variables} \\
    $\v\xi_{k, 0:T}^{(j)} \leftarrow \ p(\v\xi_{k, 0:T}\vert \v h_{k, 0:T}^{(j)},\kappa_k^{(j-1)},\mu_k^{(j-1)})$\\
    \BlankLine
    
    \textcolor{blue}{\textbackslash\textbackslash\hspace{0.1cm}draw global mean log-volatility} \\
    $\mu_k^{(j)} \leftarrow \ p(\mu_k \vert \v h_{k,1:T}^{(j)}, \kappa_k^{(j-1)}, \v\xi_{k, 1:T}^{(j)})$\\
    \BlankLine
    
    \textcolor{blue}{\textbackslash\textbackslash\hspace{0.1cm}draw global log-volatility persistence} \\
    $\kappa_k^{(j)} \leftarrow \ p(\kappa_k \vert \v h_{k, 1:T}^{(j)}, \mu_k^{(j)}, \v\xi_{k, 1:T}^{(j)})$\\
    \BlankLine
    }

     \textcolor{blue}{\textbackslash\textbackslash\hspace{0.1cm}draw pre-sample observations } \\
    $\v x{^\top_{0}}^{(j)} \leftarrow \ p(\v x{^\top_{0}} \vert \v\theta^{(j)}_{0:T},\sigma{^{2(j-1)}_{{0}:T}},\varepsilon^{(j-1)}_{\text{-}{(q_{\max}-1)}:T}, y_{1:T}))$\\
    \BlankLine

         \textcolor{blue}{\textbackslash\textbackslash\hspace{0.1cm}draw the error state vector } \\
    $\v\varepsilon^{(j)}_{\text{-}{(q_{\max}-1)}:T} \leftarrow \ p(\v\varepsilon_{\text{-}{(q_{\max}-1)}:T} \vert \v\theta^{(j)}_{0:T},\sigma{^{2(j-1)}_{{0}:T}}, \v x_0^{\top(j-1)}, y_{1:T})$\\
    \BlankLine

         \textcolor{blue}{\textbackslash\textbackslash\hspace{0.1cm}draw the error standard deviations } \\
    $\sigma_{1:T}^{2(j)} \leftarrow \ p(\sigma^{2}_{1:T} \vert \v\varepsilon_{\text{-}{(q_{\max}-1)}:T})$\\
    \BlankLine
} 
\KwOutput {
draws from the joint posterior of $\v\theta_{0:T},\v h_{0:T},\sigma_{1:T}^2,\v{\varepsilon}_{\text{-}{(q_{\max}-1)}:T}, \v{x}_{0}^\top, \v\mu$ and $\v\kappa$.
}

\BlankLine
\caption{Gibbs sampling from joint posterior $p(\v\theta_{0:T},\v h_{0:T},{\varepsilon}_{\text{-}q_{\text{-}{(q_{\max}-1)}:T}}, {\v x_0^T},\sigma_{1:T}^2, \v\mu,\v\kappa \vert y_{1:T})$ \label{alg:gibbs_algorithm}}
\end{algorithm}

\subsection{Updating the error paths}\label{subsec:update_errors}
The joint conditional posterior of the error paths \( p(\varepsilon_{-q_{\max}:T} | y_{1:T}, \cdot) \), where the dot \( \cdot \) serves as a placeholder for all other parameters in the model, can be sampled by rewriting the model in  \eqref{eq: SARMA_DSP} in an alternative state-space form 
\citep{hamilton2020time}:
\begin{align}
\label{eq: MA_Hamilton}
   y_t &= \mathbf{x}_t^\top \tilde{\boldsymbol{\phi}}_t + \mathbf{f}^\top \boldsymbol{\alpha}_t\\
   \boldsymbol{\alpha}_{t} &= \mathbf{G} \boldsymbol{\alpha}_{t-1} + \boldsymbol{\omega}_{t},
\end{align}
where $\v{\alpha}_t = (\varepsilon_t, \ldots, \varepsilon_{t-(q_{\max}+1)})$ represents the unknown state vector, \(\boldsymbol{\alpha}_t \in \mathbb{R}^{(q_{\max}+1)}\). The state transition matrix  is
\[
\mathbf{G} =
\begin{pmatrix}
\*0^\top_{q_{\max}} & 0 \\
\*I_{q_{\max}} & \*0_{q_{\max}} 
\end{pmatrix}
\]
where $\*0_{k}$ is a $k\times 1$ zero vector. The vector  \(\mathbf{f} = (1, \tilde{\psi}_{t1}, \ldots, \tilde{\psi}_{tq_{\max}})^\top\), comprises all $q_{\max}$  MA coefficients, including the zero coefficients obtained after the polynomial multiplication in \eqref{eq: dynamic_ARMA_multiplied_out}. The vector $\tilde{\boldsymbol{\phi}}_t$ contains the non-zero AR coefficients in  \eqref{eq: dynamic_ARMA_multiplied_out}. The vector $\boldsymbol{\omega}_t = (\varepsilon_t, 0, \ldots, 0)^\top$ is $q_{\max}+1$ dimensional. 

Given \(\tilde{\boldsymbol{\phi}}\), \(\tilde{\boldsymbol{\psi}}\), and \(\v x_t^\top\), the model \eqref{eq: MA_Hamilton} is conditionally linear Gaussian, and the joint conditional posterior of the errors \(p(\v{\varepsilon}_{-q_{\max}:T} | y_{1:T}, \cdot)\) can be directly sampled using the FFBS algorithm.  The sampling can be accelerated by noting that the  observation equation \eqref{eq: MA_Hamilton} does not contain an error term and the filter can be terminated at an earlier time $t^{*}<T$  when the posterior state covariance matrix degenerates to zero \citep{chib1994bayes}; the remaining $\varepsilon_t$ for $t>t^{*}$ can then be computed recursively from the sampled $\v{\varepsilon}_{-q_{\max}:t^{*}}$ path. 

The algorithm requires a Gaussian prior for the initial state, \( \boldsymbol{\alpha}_0 \). Following \cite{marriott1993bayesian}, we choose an informative, data-driven prior for \( \boldsymbol{\alpha}_0\), specifically \( \boldsymbol{\alpha}_0 \sim N(\boldsymbol{0}, \hat{\sigma}^2\boldsymbol{I}_{q_{\max}+1}) \), where \( \hat{\sigma}^2 \) is the residual variance estimated from fitting the static ARMA model on the first 30 observations. See \cite{marriott1993bayesian} for a discussion of potential challenges associated with choosing uninformative priors in ARMA models, and \cite{cox2023sparse}  for time-varying models in general.

\subsection{Updating the pre-sample observations }\label{subsec:update_y0}
The joint conditional posterior for the pre-sample lags $\boldsymbol{x}_0=(y_0,y_{-1},\ldots,y_{-p_{\max}+1})^\top$ is sampled using a similar approach as in Section \ref{subsec:update_errors}. The model in  \eqref{eq: SARMA_DSP} is now recast as the state-space model  \citep{hamilton2020time}

\begin{align}
\label{eq: AR_Hamilton}
   y_t &= \mathbf{z}_t^\top \tilde{\boldsymbol{\psi}}_t + \mathbf{f}^\top \boldsymbol{\alpha}_t \\
   \boldsymbol{\alpha}_{t} &= \mathbf{G_t} \boldsymbol{\alpha}_{t-1} + \boldsymbol{\omega}_{t},
\end{align}
where the state is now \(\boldsymbol{\alpha}_t = (y_t, \ldots, y_{t-(p_{\max}-1)})\), and
\[
\mathbf{G}_t =
\begin{pmatrix}
\tilde{\phi_{1t}} & \tilde{\phi}_{2t} & \ldots & \tilde{\phi}_{p_{\max}-1,t} &  \tilde{\phi}_{p_{\max,t}}  \\
1 & 0 & 0 & \ldots & 0 \\
0 & 1 & 1 & \ldots & 0 \\
\vdots & \vdots & \vdots & \ddots & \vdots \\
0 & 0 & \ldots & 1 & 0 \\
\end{pmatrix}
\]
where the first row is given by all $p_{\max}$ AR coefficients in \eqref{eq: dynamic_ARMA_multiplied_out}, including the exact zeros. The vector of regressors \( \mathbf{z}_t \) contains the lagged values of the errors \( \varepsilon_t \), and \( \mathbf{f} = (1, 0, \ldots, 0) ^\top\). Finally, $\tilde{\boldsymbol{\psi}}_t$ are the non-zero MA coefficients  in \eqref{eq: dynamic_ARMA_multiplied_out} and $\v {\omega}_t = (\varepsilon_t, 0, \ldots, 0)^\top$.

Given \(\tilde{\boldsymbol{\phi}}\), \(\tilde{\boldsymbol{\psi}}\), and \(\v z_t\), the model in  \eqref{eq: AR_Hamilton} is conditionally linear Gaussian. Consequently, the joint conditional posterior \(p(\v \alpha_t | y_{1:T}, \cdot)\), and thus \(p(\v x_0 | y_{1:T}, \cdot)\), can be sampled directly  using the FFBS algorithm. Since the observations \(\v y_t\) are fully known for \(t=1,\ldots, T\) we only need to run the filter up to time $p_{\max}$.

The initial prior for \(\boldsymbol{\alpha}_0\) is chosen in a data-driven way following \cite{marriott1993bayesian}. Specifically, we use \(\boldsymbol{\alpha}_0 \sim N(\boldsymbol{0}, \hat{\v \gamma}_0)\), where \(\hat{\gamma}_0\) is the estimated variance of the observed \(y_t\). Given the implied non-stationary nature of the observed data \(y_t\), we estimate the variance from the first $30$ observations of the series.

\subsection{Updating the noise process}\label{subsec: update_noise}
We extend the SV(1) model of \cite{kim1998stochastic} by using the following DSP prior
\begin{align}
\label{eq: noise_process}
\varepsilon_t & \sim N(0, e^{h_t^* })\nonumber \\
{h_t^* } & = {h_{t-1}^* } + \nu_{t}^* , \hspace{3.5cm} \nu_{t}^*  \overset{\mathrm{ind}}{\sim} N\big(0,  e^{h_t })\big)  \nonumber \\ 
    h_{t} &= \mu + \kappa(h_{t-1}-\mu)+\eta_{t}, \hspace{1.95cm}   
    {\eta}_{t} \overset{\mathrm{iid}}{\sim} {Z}(1/2,1/2, 0, 1).
\end{align}
The global-local properties of the DSP prior allows the volatility to both remain constant during long spells and to change rapidly or even jump at other times. 

Both $h^{*}_{0:T}$ and $h_{0:T}$ can be efficiently sampled by following the algorithm of \citet{kastner2014ancillarity} and \citet{kastner2019dealing}. In practice, we follow the framework of \citet{kowal2019dynamic}, developed for models with the DSP priors on the log-innovation variance.

\section{Simulation experiments}\label{sec:simulations}
This section evaluates the method's ability to capture temporal dynamics of varying complexity, including both abrupt and gradual changes. Section \ref{subsec:setup_sim} describes the data-generating processes, simulation settings and the benchmark model, while Section \ref{subsec:perfomance_metrics} outlines the evaluation metrics. Sections \ref{subsec:experiment1} and \ref{subsec:experiment2} present the two simulation experiments: a TV-ARMA(1,1) model illustrating, among other aspects, the near-root-cancellation effects, and a single seasonal TV-SMA$(1,1)_{12}$ model. 

\subsection{Setup and benchmark models}\label{subsec:setup_sim}
We simulate $50$ time series from the data generating process for each experiment. The simulated time series consist of $T=500$ observations, and we perform \num{10000} draws from the posterior distribution for each method, after a burn-in of \num{3000} iterations. The post burn-in draws are thinned by a factor of $10$, i.e., leaving \num{1000} thinned draws for inference on the spectral density. In all experiments we assume that the orders of the AR and MA components, $ p_{\max}$ and $ q_{\max}$ are known. Experiment 3 in \citet{fagerberg2026time} shows for the pure AR case that the DSP prior can handle mis-specified models with redundant lags quite well.

There is a lack of readily available packages for estimating time-varying ARMA models, so we restrict the comparison of the proposed TV-SARMA to the AdaptSpec method \citep{rosen2012adaptspec, BertolucciBayesSpec2021}, a semiparametric Bayesian model which is often used for benchmarking and is efficiently implemented in the R/C++ package \texttt{BayesSpec}. AdaptSpec estimates a local spectral density using smoothing splines over time segments sampled by the algorithm using a reversible jump MCMC algorithm. The default settings for AdaptSpec are used for Experiment 1 (TV-ARMA model), but we increase the default number of spline knots from 10 to 15 for Experiment 2 (TV-SMA model). This is to ensure a fair comparison, since AdaptSpec was not specifically designed for seasonal models.  Preliminary experiments indicate that this adjustment improves AdaptSpec's  ability to capture the highly multimodal spectral density in the seasonal model.

All the experiments assume a conjugate Inverse-Gamma prior for the static noise variance, $\sigma^2 \sim IG(0.01, 0.01)$.  The initial error trajectory $\varepsilon_{-q:T}$ is approximated by setting $\v z^\top_0=(-\varepsilon_{-q}, \ldots, \varepsilon_0)^\top=(0,0,\ldots,0)^\top$. For the remaining $\varepsilon_{1:T}$, we use posterior prediction errors obtained as a by-product of the EKF algorithm (applied only during the initial run). We use the Gaussian prior discussed in \cite{fagerberg2026time} for the initial state of the AR and MA coefficients in the FFBSx algorithm. To ensure numerical stability, we scale the prior variance by a factor of one-half, preventing the algorithm from approaching the parameter boundaries. In Experiment 1, we terminate the FFBS filter used to infer the error state at $t=8$, and in Experiment 2 at $t=20$, which is enough for the posterior covariance matrices to be essentially a zero matrix. In the update for $\v h_{0:T}$, we use an offset equal to the machine precision in R. For a detailed discussion on the impact of the offset term on the model performance and parameter estimation, see  \citet{fagerberg2026time}.

\subsection{Performance metric}\label{subsec:perfomance_metrics}
Since AdaptSpec is defined in the frequency domain, we compare the methods based on their ability to estimate the time-varying spectral density. The spectral density of a single-seasonal ARMA process at time $t$ is given by \citep{wei2019time}
\begin{equation}
    \label{eq: spectral_density}
    f(t,\omega)= \frac{\sigma^2}{\pi}\frac{|\psi_{qt}(e^{-\im \omega})|^2}{|\phi_{pt}(e^{-\im \omega})|^2}\frac{|\Psi_{Qt}(e^{-\im \omega})|^2}{|\Phi_{Pt}(e^{-\im s\omega    })|^2  } \text{ for } \omega \in (0, \pi ) \text{ and } t=1,\dots,T,
\end{equation}
where $\phi_{pt}$, $\Phi_{Pt}$  and  $\psi_{pt}$, $\Psi_{Qt}$ are the usual regular and seasonal AR and MA lag polynomials, respectively, at time $t$, and $\omega \in (0, \pi ]$ is the radial frequency. This expression naturally extends to  models with multiple seasonal periods. Following \citet{rosen2012adaptspec}, we use the Mean Square Errors (MSE) of the estimated log spectral density over all time periods as the performance metric
\begin{equation}
     \label{eq: MSE}
     \mathrm{MSE}( \widehat{\log f}) =\frac{1}{T m}\sum_{t=1}^{T}\sum_{k=1}^{m}\big(\widehat{\log f(t,\omega_k)}-\log f(t,\omega_k)\big)^2,
 \end{equation}
where $\log f(t,\omega)$ is the true log spectral density at time $t$ and frequency $\omega$, and $\widehat{\log f(t,\omega)}$ is its estimate, and $m=100$ is the total number of frequencies on a grid from $0$ to $\pi$. 

The posterior median and 95\% highest posterior density intervals (HDIs) are used as summaries for the estimated parameters. The posterior median is used since it is more robust to outliers in the MCMC.

\subsection{Experiment 1 -  Non-Seasonal TV-ARMA }\label{subsec:experiment1}
\begin{figure} 
 \centering
 \includegraphics[width=0.8\textwidth]{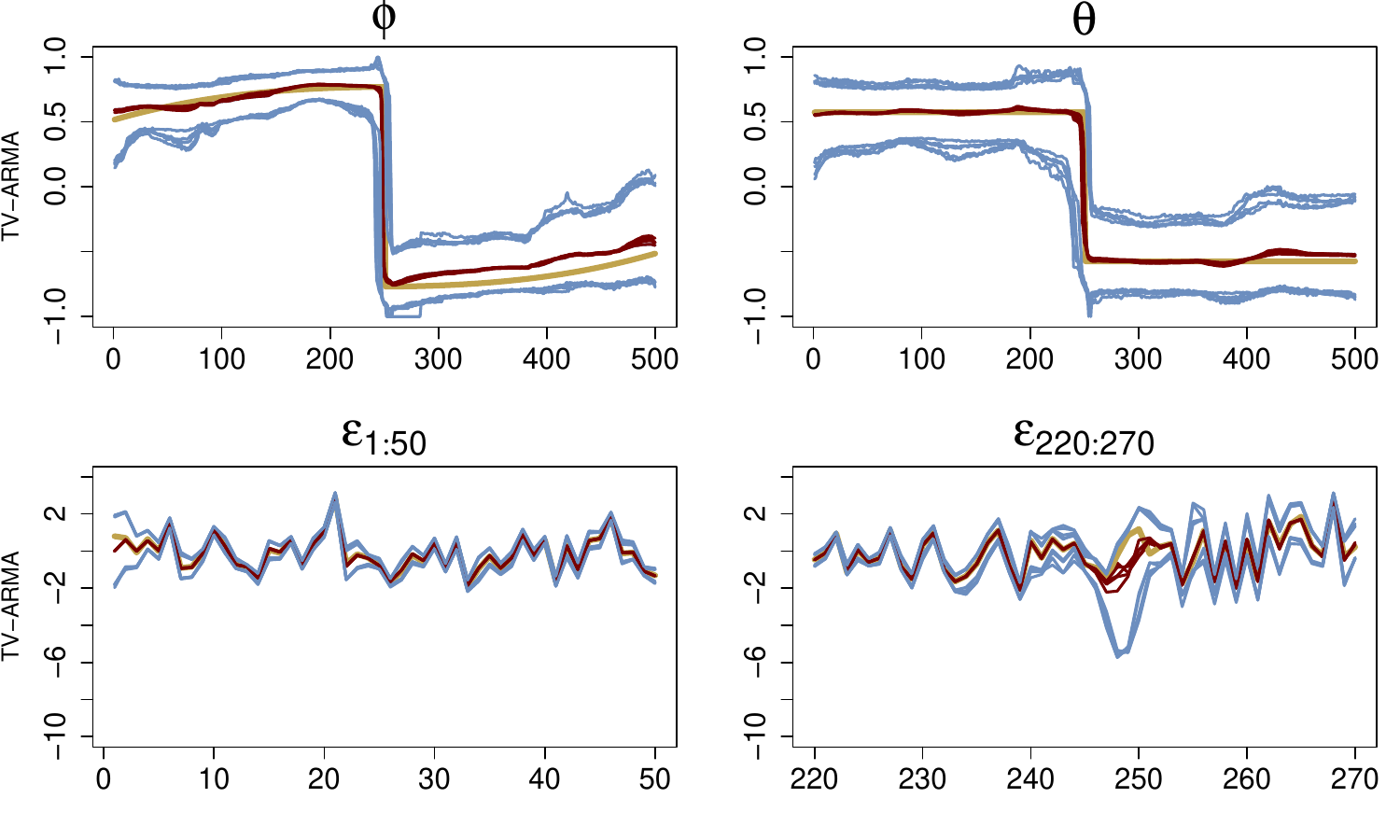}

   \caption{Experiment 1. Assessing MCMC convergence of FFBSx algorithm by re-estimating the model from five different seeds using five different initial values for $\v h_{1:T}$. All runs are for the same dataset. The red and blue lines are posterior medians and $95\%$ HDIs over time for each of the five repeated runs. The beige line denotes the true parameter evolution. The posterior error trajectory,  $\v \varepsilon_{-q_{\max}:T}$, is displayed for two time segments,  $t=1,\ldots,50$  and  $t=220,\ldots,270$ for visual clarity. }\label{fig:exp1convergence}
\end{figure}

This section demonstrates the ability of our method to track parameter evolution over time in pure ARMA models with constant noise variance. The data generating process is the time-varying ARMA$(1, 1)$
\begin{align}\label{eq: exp1}
   (1-\phi_{t} L) y_t &= (1+\psi_{t} L)\varepsilon_t.
\end{align}
with $\varepsilon_t \sim N(0,1)$. The time evolution of the unrestricted AR parameter is given by
\begin{equation}\label{eq: exp1_AR}
    \theta_{1t} = 
    \begin{cases}
       \phantom{-}0.6+0.6\sin{(\frac{\pi t}{T})} & \text{ for } t=1,\ldots,250 \\
    -0.6-0.6\sin{(\frac{\pi t}{T})}& \text{ for } t=251,\ldots,500 
    \end{cases}
\end{equation}
while the unrestricted MA parameter follows
\begin{equation}\label{eq: exp1_seas1}
    \theta_{2t} =     
    \begin{cases}
       \phantom{-}0.7 & \text{ for } t=1,\ldots,250 \\
    -0.7& \text{ for } t=251,\ldots,500 
    \end{cases}
\end{equation}
The beige lines in Figure~\ref{fig:exp1convergence} show the evolutions of the regular AR and MA parameters, $\phi_{t}$ and $\psi_{t}$,  respectively. 

We first explore the numerical performance of the Gibbs sampler. Figure \ref{fig:exp1convergence} assesses the convergence of the algorithms by plotting the posterior medians and equal-tail $95\%$ HDIs intervals for the AR and MA parameters, $\phi_{1:T}$ and $\psi_{1:T}$, and the error trajectory  $\v \varepsilon_{-q_{\max}:T}$ . These results are based on five different runs on the same dataset, each with different initial values for the log-volatilities $h_{k,t}$.  The initial values for $h_{k,t}$ in the five runs are set to $-17,-16, -15,-14$ and $-13$, respectively, for all $k$ and $t$. 

\begin{figure} 
 \centering
 \includegraphics[width=0.6\textwidth]{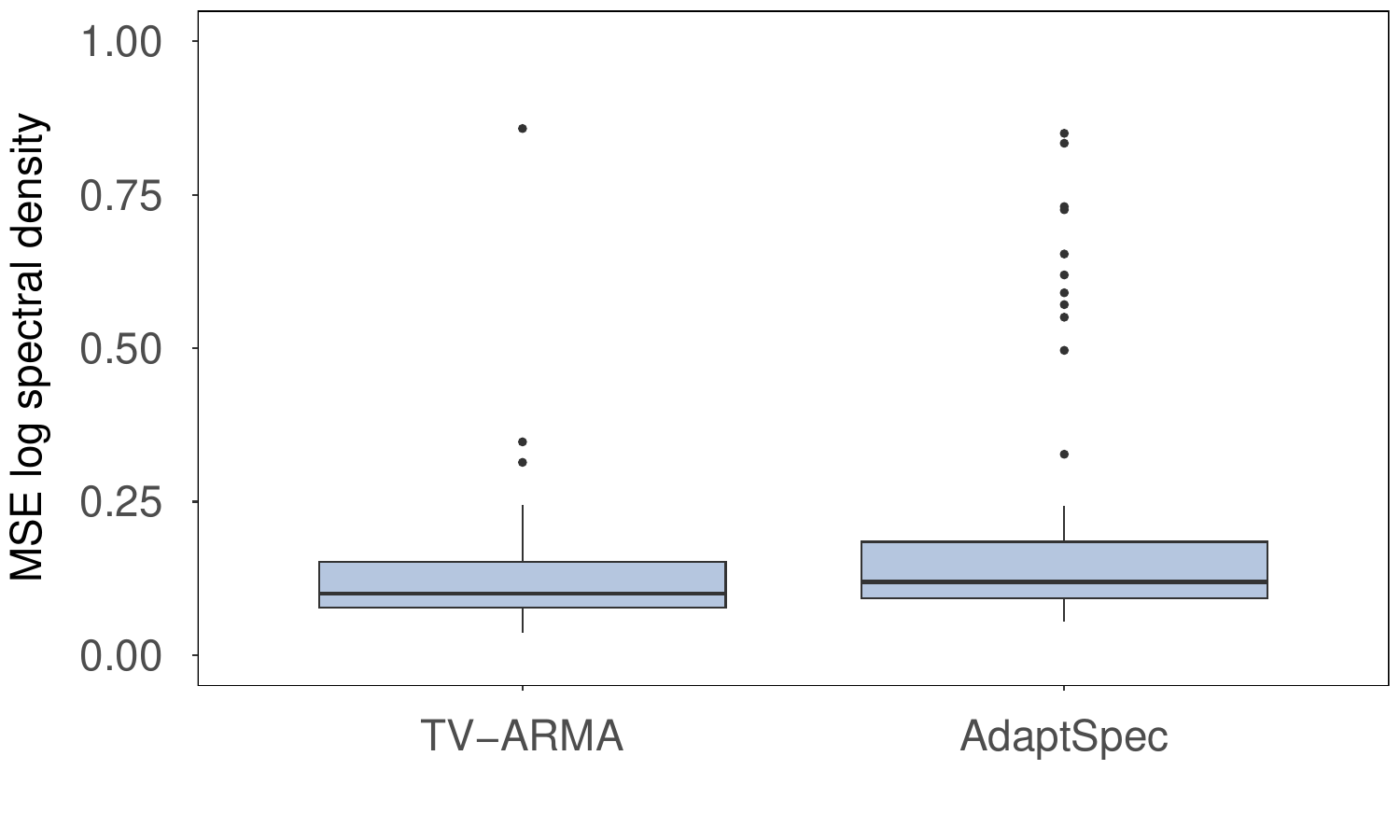}
   \caption{Experiment 1 . Box plots for the MSE for the log spectral density over time and frequency from 50 simulated datasets.   }\label{fig:exp1boxplots}
\end{figure}

Figure~\ref{fig:exp1convergence} shows that the algorithm tracks the evolution of the AR and MA parameters well over time, including the rapid shift. It also correctly infers the errors, $\varepsilon_{1:T}$. In the vicinity of time period $t=250$, where the parameters undergo a rapid change, the credible intervals for the errors grow wider. This is likely due to the natural interplay between the errors and the MA parameters which can complicate identification, especially during abrupt changes. The simultaneous estimation with the AR parameters may exacerbate this issue, where problems of near root cancellation may be lurking. The effect of root cancellation is further explored below. The posterior result across the five runs in Figure~\ref{fig:exp1convergence} are similar, indicating satisfactory convergence of the Gibbs sampler.

\begin{figure} 
 \centering
 \includegraphics[width=1\textwidth]{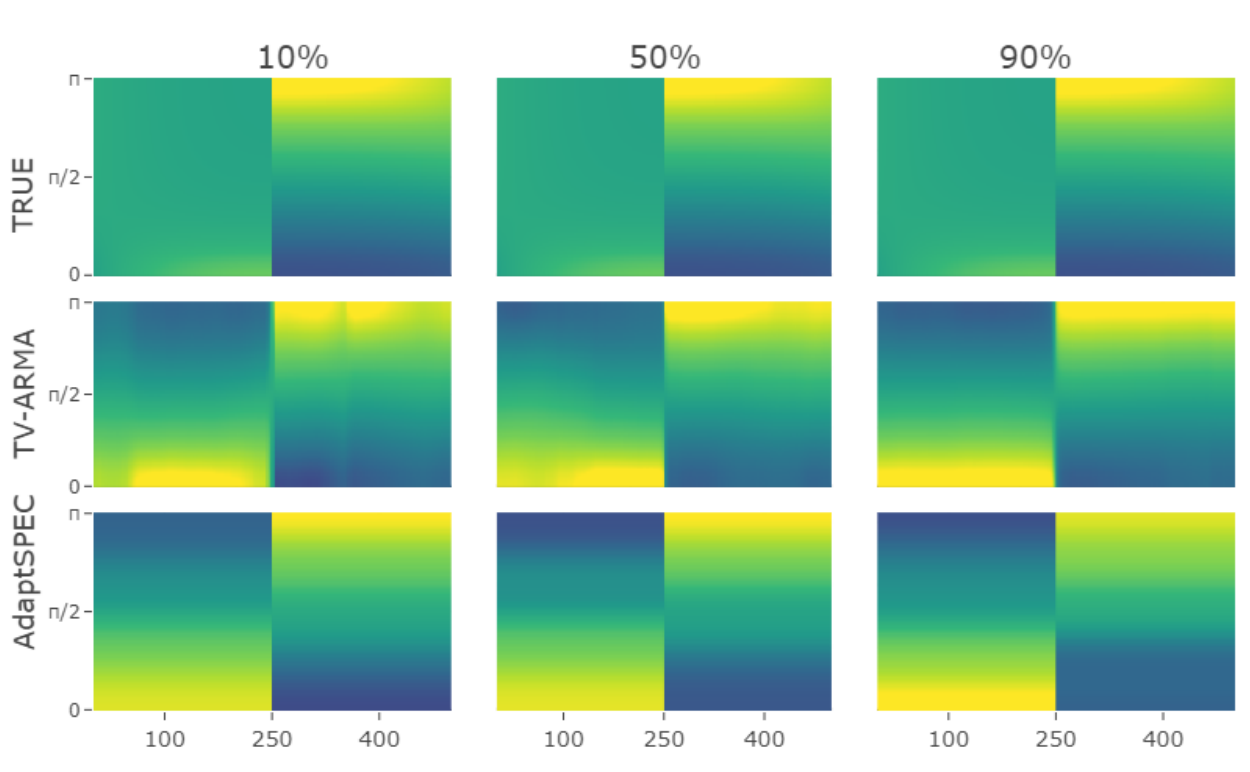}
 
 \caption{Experiment 1. Heatmaps of the estimated log spectral density over time for the different methods. The columns correspond to different datasets chosen from the percentiles of each model’s MSE distribution, to show the performance of each model when it performs well (10\%), average (50\%) and poorly (90\%). }\label{fig:exp1heatmaps}
\end{figure}

Figure \ref{fig:exp1boxplots} presents box plots of the MSE metric in \eqref{eq: MSE} for the compared models across 50 simulated datasets. The TV-SARMA performs slightly better than AdaptSpec in terms of median and interquartile range, but the main difference is that AdaptSpec model exhibits a much higher number of sizeable outliers, indicating greater variability and less robustness than the TV-SARMA model. 

Figure~\ref{fig:exp1heatmaps} displays heatmaps of the estimated time-varying log spectral density for the compared models.  Following \cite{rosen2012adaptspec}, we plot the posterior median of the time varying log spectral densities for three different datasets, corresponding to the 10\%, 50\%, and 90\% percentiles of the MSE values for each method. This illustrates  each method's performance when it is among its best (10\%), average (50\%), and worst (90\%). Note that the columns in Figure~\ref{fig:exp1heatmaps} are therefore for potentially different datasets for different methods. Figure  \ref{fig:exp1spectra_time} in the supplementary material gives an alternative view by plotting the estimated time-evolution at some selected frequencies, and Figure~\ref{fig:exp1spectra_freq} plots the fitted log spectral densities at three different time points. The plots show that both methods provide comparable results in estimating the spectral densities over time and frequencies, with the TV-SARMA model achieving a slightly better fit overall. The relatively similar performance of the two methods in this experiment is probably because the parameters in the DGP are fairly constant with one big jump, something that AdaptSpec's segmented approach can handle rather well.

\begin{figure} 
 \centering
 \includegraphics[width=0.8\textwidth]{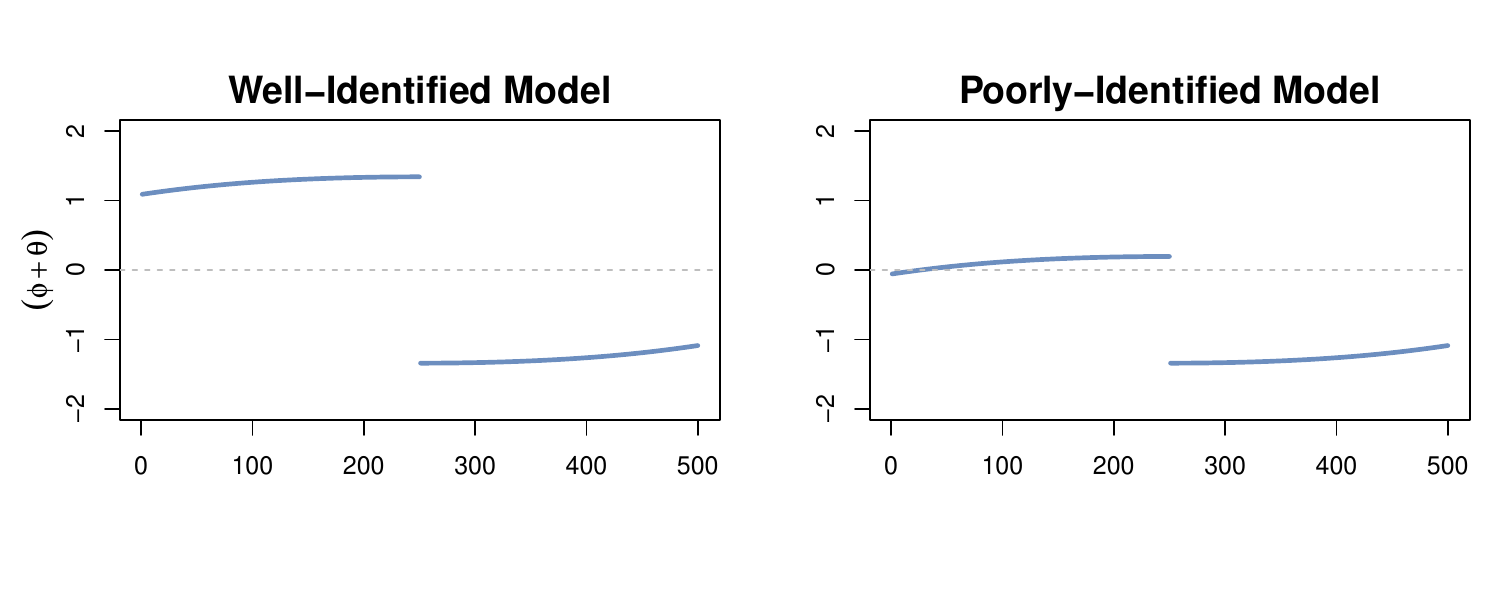}
   \caption{Experiment 1. The sum of the time-varying AR, $\phi$, and MA, $\theta$, parameters as an identification tool, illustrating the potential issue of root cancellation in an ARMA(1,1) model. The left panel shows a well-identified model where the sum remains far from zero, while the right panel depicts a poorly-identified model with periods where the sum approaches zero.}\label{fig:ident_param}
\end{figure}

\begin{figure}
 \centering
 \includegraphics[width=0.8\textwidth]{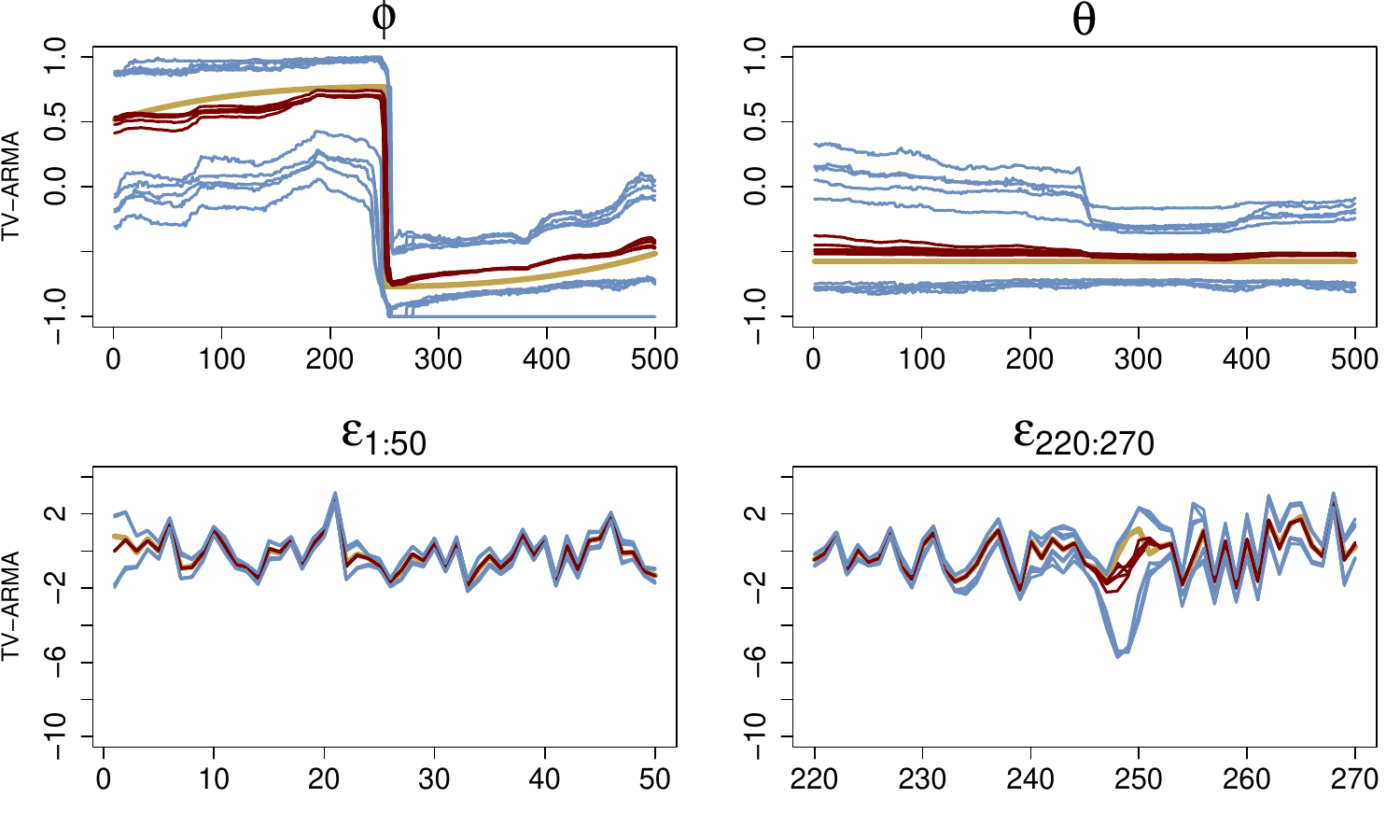}

   \caption{Experiment 1 (\textit{poorly-identified model}). Assessing MCMC convergence of FFBSx algorithm by re-estimating the model from three different seeds using three different initial values drawn for the global hyperparameter $\v\mu$. All runs are for the same dataset. The red and blue lines are posterior medians and $95\%$ HDIs over time for each of the three repeated runs. The beige line denotes the true parameter evolution. The posterior error trajectory,  $\v \varepsilon_{-q_{\max}:T}$, is displayed for two time segments,  $t=1,\ldots,50$  and  $t=220,\ldots,270$ for visual clarity. .  }\label{fig: exp1_convergence2}
\end{figure}

To illustrate the challenges posed by root cancellation, we simulate from the above ARMA(1,1) model, but with a parameter evolution where the roots nearly cancel out during certain time segments
\begin{equation}\label{eq: exp1_AR_rootcancel}
    \theta_{1t} = 
    \begin{cases}
      \phantom{-}0.6+0.6\sin{(\frac{\pi t}{T})} & \text{ for } t=1,\ldots,250 \\
    -0.6-0.6\sin{(\frac{\pi t}{T})}& \text{ for } t=251,\ldots,500 \\
    \end{cases}
\end{equation}
while the unrestricted MA parameter follows
\begin{equation}\label{eq: exp1_seas_rootcancel}
    \theta_{2t} =     -0.7 \text{ for } t=1,\ldots,500. \\
\end{equation}
With this time evolution, the ARMA parameters nearly cancel each other out at some time points in the time segment $t=1,.., 250$. In ARMA($1,1$) models, the closer the sum $\phi+\theta$ is to zero, the more severe is the root cancellation. When the sum is exactly zero, the model simplifies to pure white noise and the AR and MA parameters are unidentified. Figure \ref{fig:ident_param} illustrates the sum of the coefficients for both well-identified model in \eqref{eq: exp1_AR} and the poorly identified with near root cancellation model in \eqref{eq: exp1_AR_rootcancel}. The posterior medians in Figure \ref{fig: exp1_convergence2}, are still close to the true values of the parameters, but the HDIs over the segment $t=1,.., 250$ are wide and MCMC convergence is poor. Root cancellation is a long standing issue in the ARMA literature \citep{ kleibergen2000bayesian} that users should be aware of, especially in time-varying parameter models where the process may approach root cancellation only at certain time points. 

\subsection{Experiment 2 - Seasonal TV-SMA }\label{subsec:experiment2}
\begin{figure}
 \centering
 \includegraphics[width=0.8\textwidth]{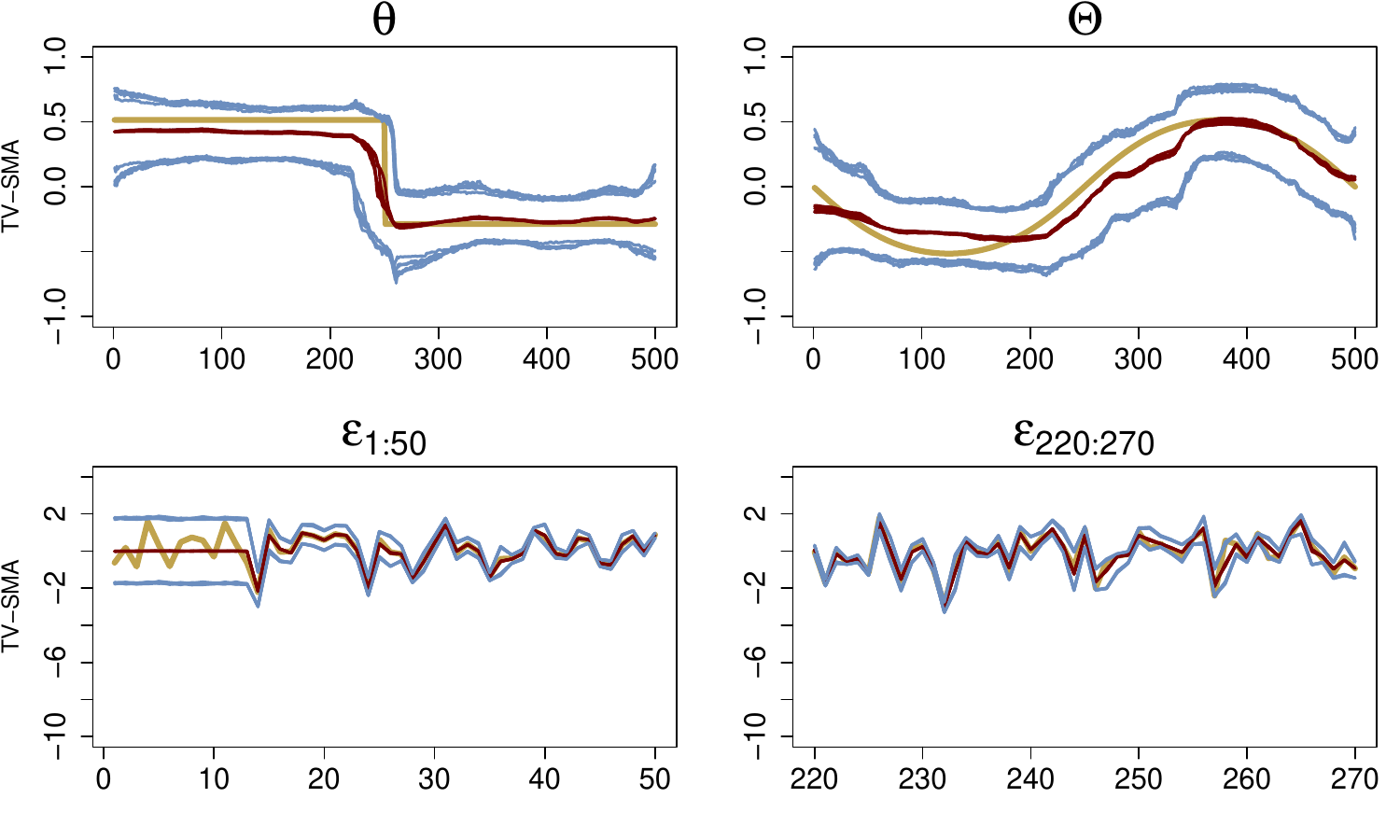}
 
   \caption{Experiment 2. Assessing MCMC convergence of FFBSx algorithm by re-estimating the model using five different seeds and five different initial values for the global hyperparameter $\v\mu$. All runs are for the same dataset. The red and blue lines are posterior medians and $95\%$ HDIs over time for each of the five repeated runs. The beige line denotes the true parameter evolution. The error trajectory  $\v \varepsilon_{-q_{\max}:T}$ is displayed for two time segments,  $t=1,\ldots,50$  and  $t=220,\ldots,270$ for visual clarity. . }\label{fig:exp2convergence}
\end{figure}

The second experiment simulates data from the seasonal time-varying moving average model, TV-SMA$(1,1)_{12}$:
\begin{align}\label{eq:exp1}
y_t &=    (1+\psi_{1t} )(1+\Psi_{1t} L^{12} )\varepsilon_t,
\end{align}
where $\varepsilon_t \sim N(0,1)$, and $t=1,\ldots, 500$. The time evolution of the unrestricted non-seasonal parameter is
\begin{align}\label{eq: Exp1_nonseas}
\theta_{1t} &=\begin{cases}
        \phantom{-}0.6\sin(\pi \frac{t}{T})& \text{ for } t=1,\ldots,250 \\
       -0.6\sin(\pi  \frac{t}{T}) & \text{ for } t=251,\ldots,500
\end{cases} \\ \nonumber 
\end{align}
and the time evolution of the unrestricted seasonal parameter is
\begin{align}\label{eq: Exp1_seas}
\theta_{2t} &=  0.6\sin(2\pi t/T) &\text{for all} \ t    
\end{align}
The beige lines in Figure~\ref{fig:exp2convergence} show the evolutions of the regular and seasonal MA parameters, $\psi_{t}$ and  $\Psi_{t}$, respectively. 

The numerical performance of the Gibbs sampler is explored first, following the same setup as in Experiment 1. Figure \ref{fig:exp2convergence} shows that the algorithm effectively tracks the evolution of the MA parameters over time,  including  rapid shifts, while accurately identifying the posterior distribution of the errors. Note the large uncertainty of the $s+1=13$ initial errors. The posterior distribution is stable across the five runs with different initial values.

\begin{figure} 
 \centering
 \includegraphics[width=0.5\textwidth]{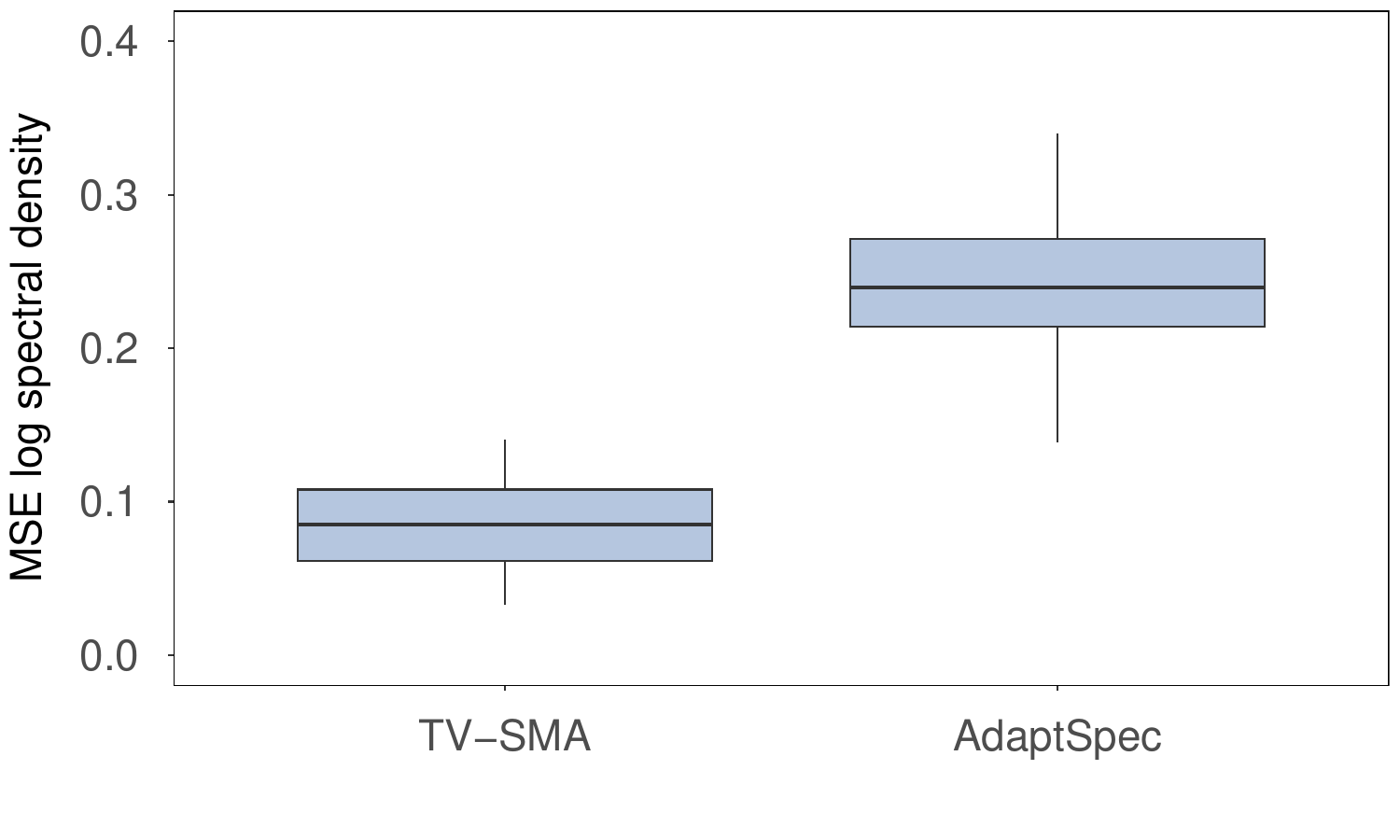}
   \caption{Experiment 2. Box plots for the MSE for the log spectral density over time and frequency from 50 simulated datasets.   }\label{fig:exp2boxplots}
\end{figure}

Figure \ref{fig:exp2boxplots} presents box plots of the MSE metric \eqref{eq: MSE} for the compared models across 50 simulated datasets. The TV-SARMA model outperforms the AdaptSpec model with a consistently low MSE across the datasets. The AdaptSpec model is not specifically tailored to handle seasonal data and its semiparametric fit using splines is not able to capture the multiple peaks in the spectral density well, despite the increased number of splines terms. 

\begin{figure}
 \centering
 \includegraphics[width=0.9\textwidth]{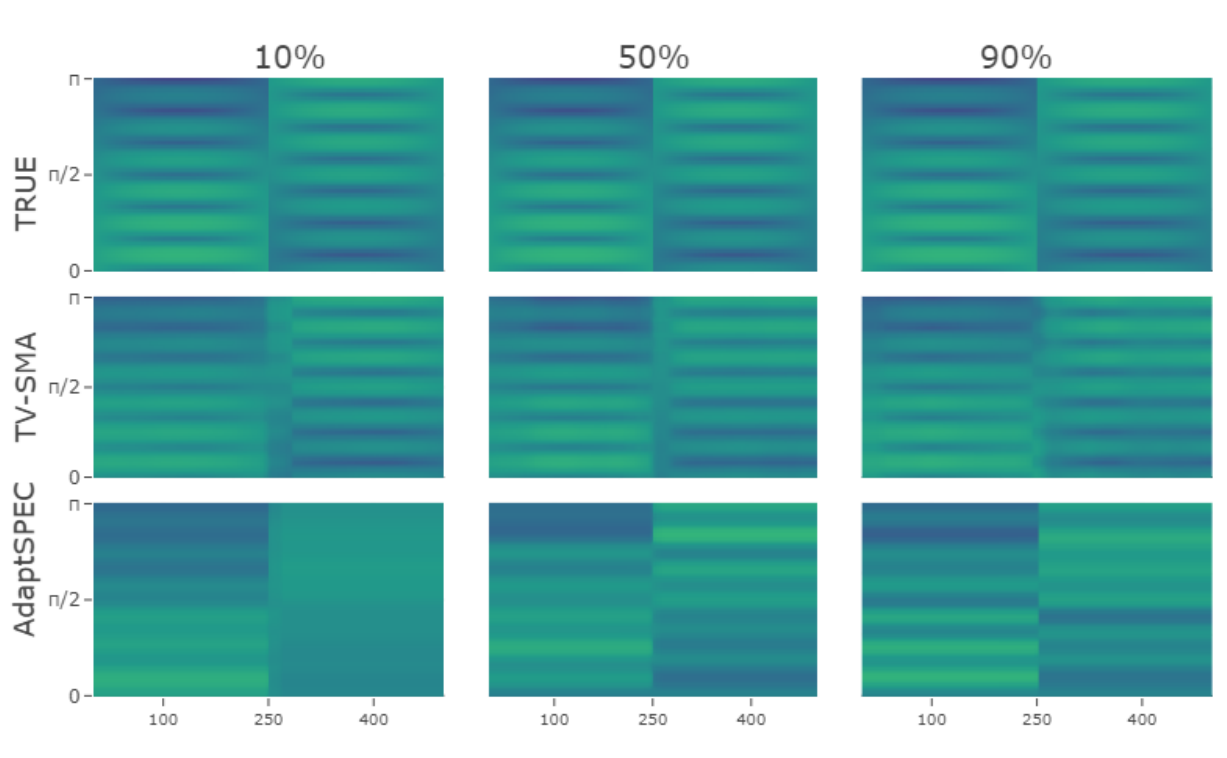}

 \caption{Experiment 2. Heatmaps of the estimated log spectral density over time for the different methods. The columns correspond to different datasets chosen from the percentiles of each model’s MSE distribution, to showcase the performance of each model when it performs well (10\%), average (50\%) and poorly (90\%). }\label{fig:exp2spectra}
 \label{fig:exp1spectra}
\end{figure}

\begin{figure} 
 \centering
 \includegraphics[width=0.9\textwidth]{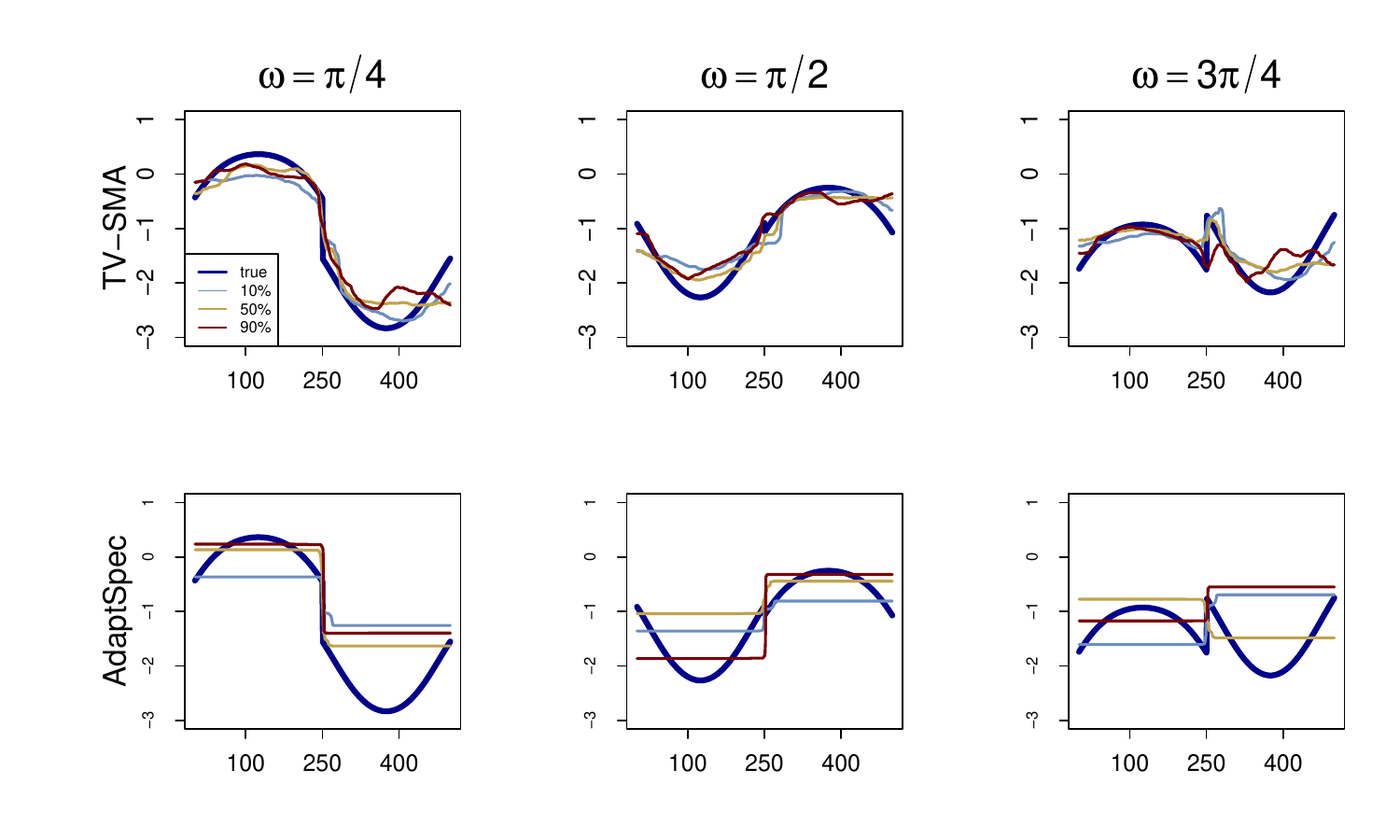}
   \caption{Experiment 2. Time evolution of the log spectral density at three different frequencies (columns) for several different model (rows). The dark blue line is the true spectral density and the three colored lines are the posterior median estimates from three different datasets chosen from the MSE percentiles.  }\label{fig:exp2spectra_time}
\end{figure}

\begin{figure} 
 \centering
 \includegraphics[width=0.9\textwidth]{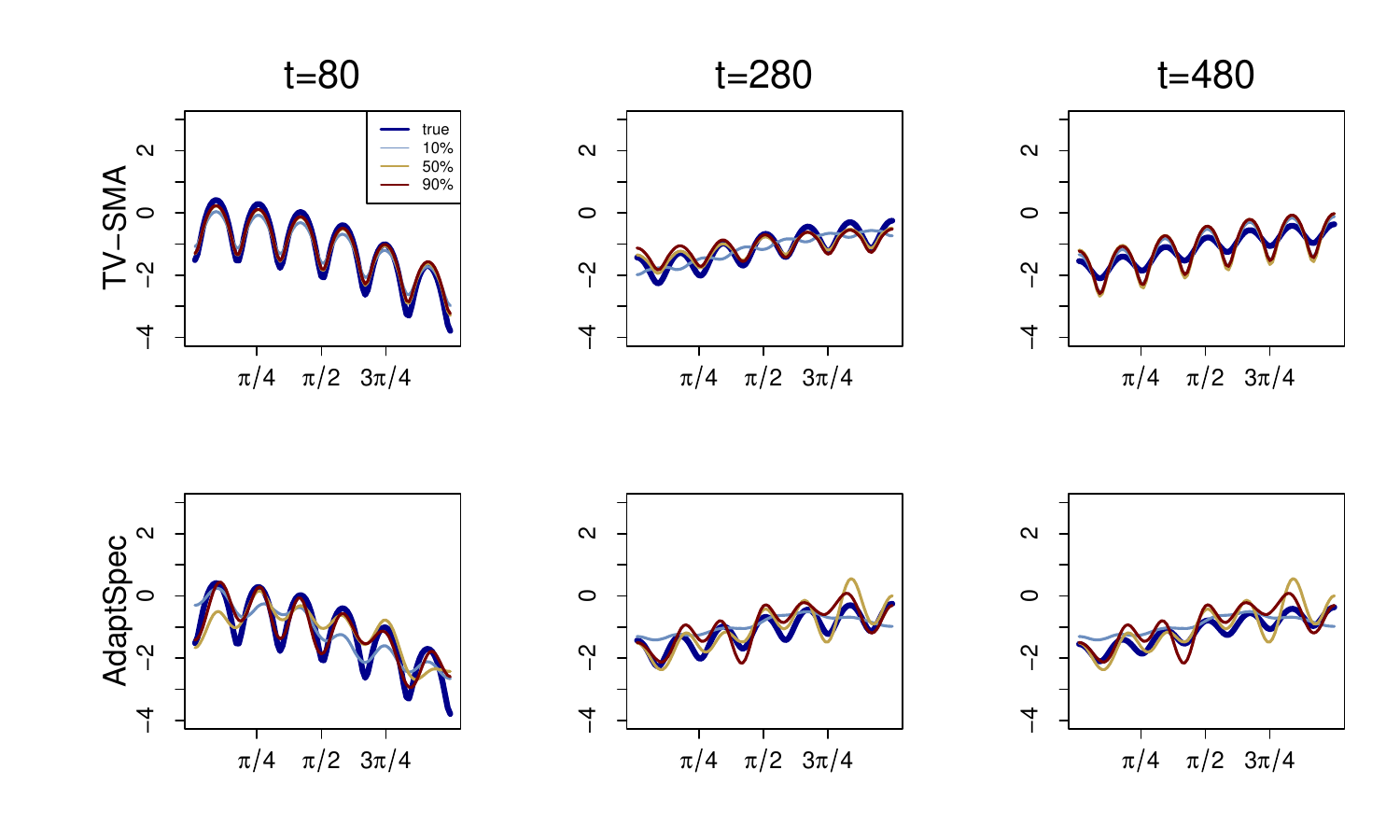}
   \caption{Experiment 2. Log spectral density at three different time points. The dark blue line is the true spectral density and the three colored lines are the posterior medians from three different datasets chosen from the MSE percentiles. }\label{fig:exp2spectra_freq}
\end{figure}

Figure \ref{fig:exp2spectra} displays heatmaps of the estimated time-varying log spectral density for the compared models.  Figure \ref{fig:exp2spectra_time} gives an alternative view by plotting the estimated time evolution at some selected frequencies. Figure \ref{fig:exp2spectra_freq} plots the fitted log spectral densities at three different time points, showing the fit for three different datasets selected from the MSE percentiles for each method.


\section{Application to US airpassenger data}\label{sec:applications}

\begin{figure}
 \centering
 \includegraphics[width=1\textwidth]{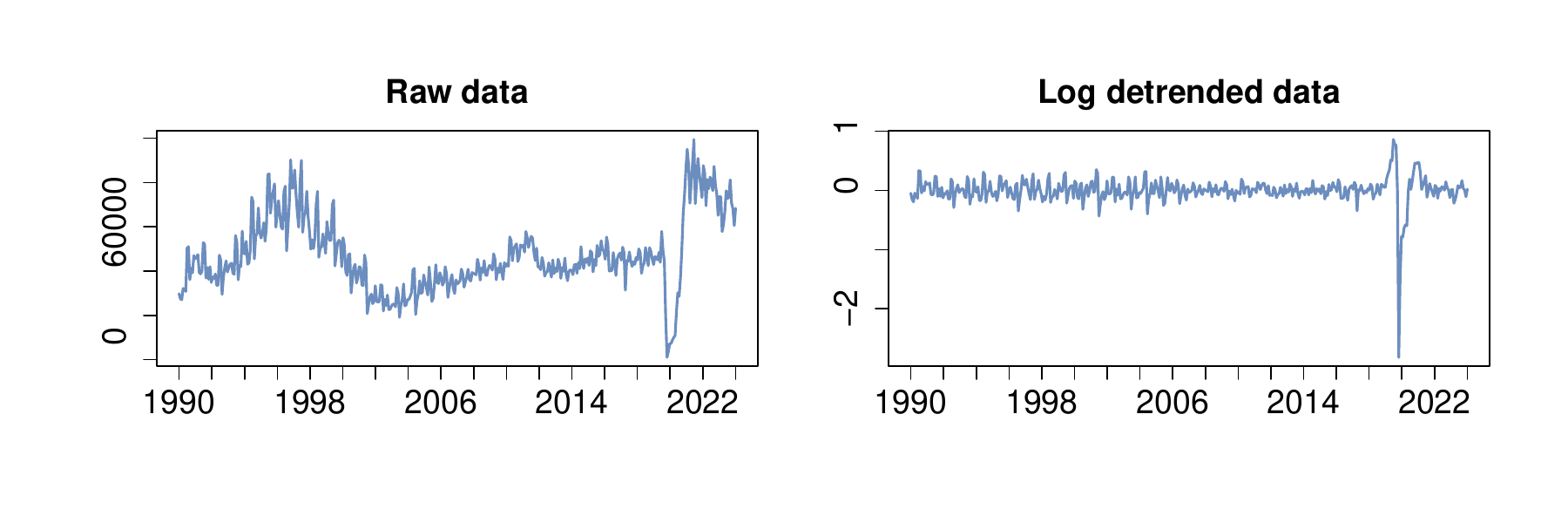}

    \caption{US air passenger data: Number of passengers traveling between JFK (New York) and Miami airports from January 1990 to July 2024. }\label{fig:usair_data}
\end{figure}

We analyze the monthly number of passengers between January 1990 and July 2024 on flights between JFK (New York) airport and Miami airport in the US; the data is taken from the Bureau of Transportation Statistics (\url{https://www.bts.gov}). To focus on the time-varying ARMA and seasonal parameters, we log-transform and detrend the data by subtracting a locally estimated mean using the \texttt{locfit} package in R, where the optimal bandwidth is selected with the \texttt{bw.ucv} function. The left panel of Figure~\ref{fig:usair_data} displays the original data, while the right panel shows the detrended data. 

We use purely seasonal MA models, as in the classic airline model of \citet{boxjen@1970}, but we do not impose differencing since our time-varying model is capable of capturing non-stationarity via the evolving parameters. 

\begin{figure}
 \centering
 \includegraphics[width=0.45\textwidth]{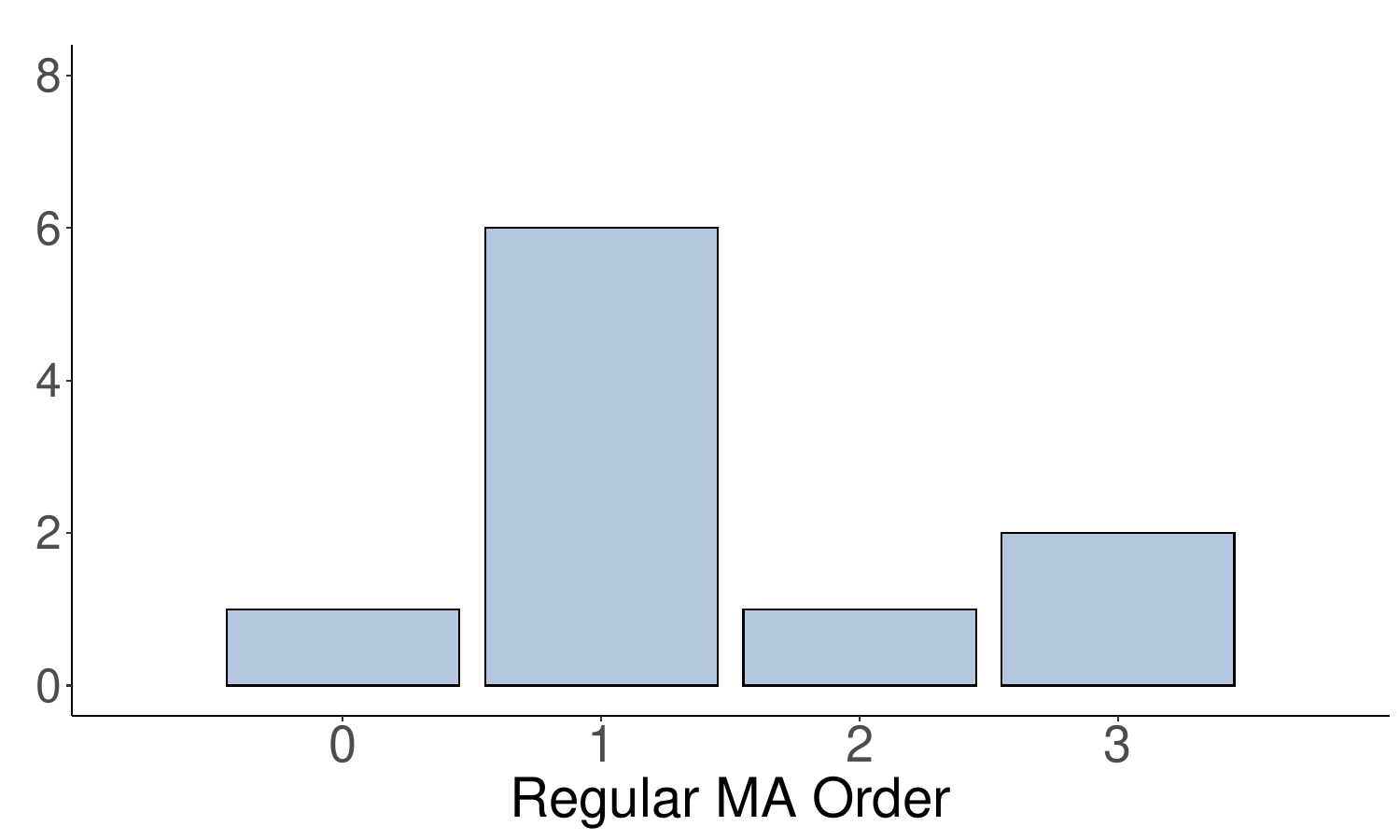}
  \includegraphics[width=0.45\textwidth]{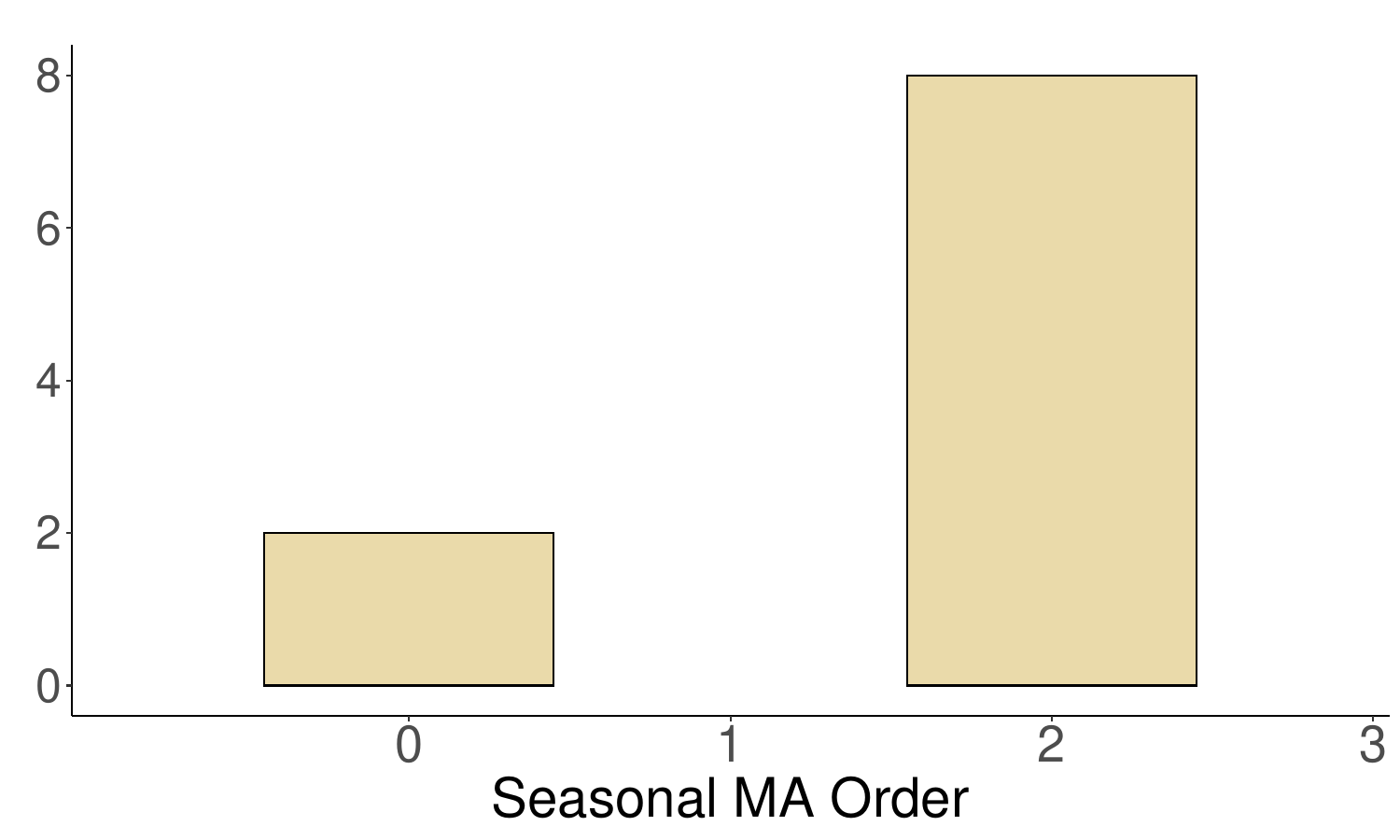}
    \caption{US air passenger data: Maximum number of regular lags (left panel) and seasonal lags (right panel) selected lags for the US airpassenger data with \texttt{auto.arima} function on segments.}\label{fig:max_order}
\end{figure}

 Given the monthly frequency of the data, we assume a seasonal period of $s = 12$. To determine the optimal number of regular and seasonal lags in TV-AR models, \cite{fagerberg2026time} use log predictive score criterion, which evaluates model performance out-of-sample. While effective, this approach can be computationally intensive since the Gibbs sampler must be run on an incrementally extended set of training observations for each observation in the test set. In this application, we use a more practically feasible rolling window method for model order selection. A seven year window is rolled across the dataset with a step size of three years of data. Within each window, the best-fitting model is identified by \texttt{auto.arima} function in the \texttt{forecast} package in R. 
Figure~\ref{fig:max_order} shows that \texttt{auto.arima} selected a maximum of three regular MA terms and two seasonal MA terms across all segments, suggesting that a TV-SMA$(3,2)_{12}$ model is the largest model to be entertained. However, the regular lag of three was only selected in the time segments spanning both the pre-pandemic and Covid-19 periods. These periods are clearly non-stationary, so models with time invariant parameters are likely to show a spurious large number of lags, compared to the model with time-varying parameters.  This suggests the model TV-SMA$(2,2)_{12}$ may be sufficient when parameters are allowed to change in every time period. When fitting the TV-SMA$(2,2)_{12}$ model to the data, the coefficient for the second lag in TV-SMA(2,2) model was consistently close to zero across all periods; see the results in Section \ref{app:usair_stability}. Consequently, we fit the simpler TV-SMA$(1,2)_{12}$ model to the data following the parsimony principle.

We model the time-varying variance of the data using both the SV($1$) model with DSP priors in Section \ref{subsec: update_noise}, and the SV(1) model of \cite{kim1998stochastic}, as implemented in the \texttt{stochvol} package in R. The models are denoted as TV-SMA (SV-DSP) and TV-SMA (SV), respectively.

\begin{figure} 
 \centering
 \includegraphics[width=1\textwidth]{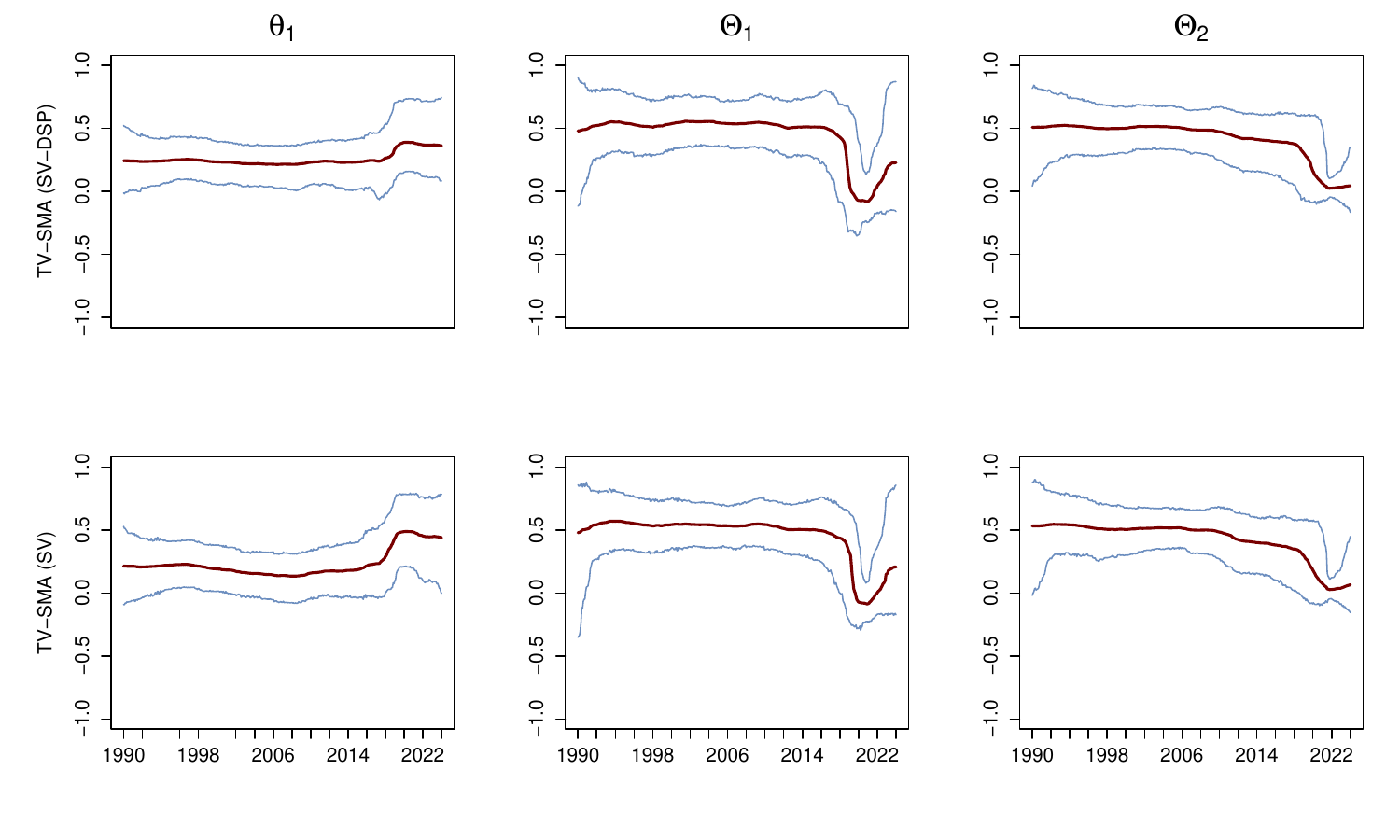}
   \caption{US air passenger data: The red and blue lines are posterior medians and $95\%$ HDIs over time for both TV-SMA$(1,2)_{12}$ (SV-DSP) (top panel) and TV-SMA$(1,2)_{12}$ (SV) (bottom panel). }\label{fig:usair_sma12_coeff}
\end{figure}

Figure \ref{fig:usair_sma12_coeff} shows the posterior medians and 95\% HDIs for the MA coefficients for the TV-SMA$(1,2)_{12}$ model from 10000 post burn-in draws for both TV-SMA (SV-DSP)  and TV-SMA (SV); Section \ref{app:usair_stability} shows that the results are stable across repeated runs. Both the regular and seasonal MA coefficients are essentially constant during the pre-pandemic period. However, during the pandemic period, the model indicates an upward shift in the regular MA coefficient, suggesting that recent shocks have a greater influence on the response variable compared to the pre-pandemic period. However, the most striking result in Figure \ref{fig:usair_sma12_coeff} is the substantially smaller seasonal variation during the pandemic, with both seasonal coefficients approaching zero. Post-pandemic, the first seasonal coefficient appears to be returning to pre-pandemic levels, while the second remains close to zero. However, given the limited availability of post-pandemic data (approximately one year), further observations are required to confirm these trends.

\begin{figure} 
 \centering
 \includegraphics[width=0.8\textwidth]{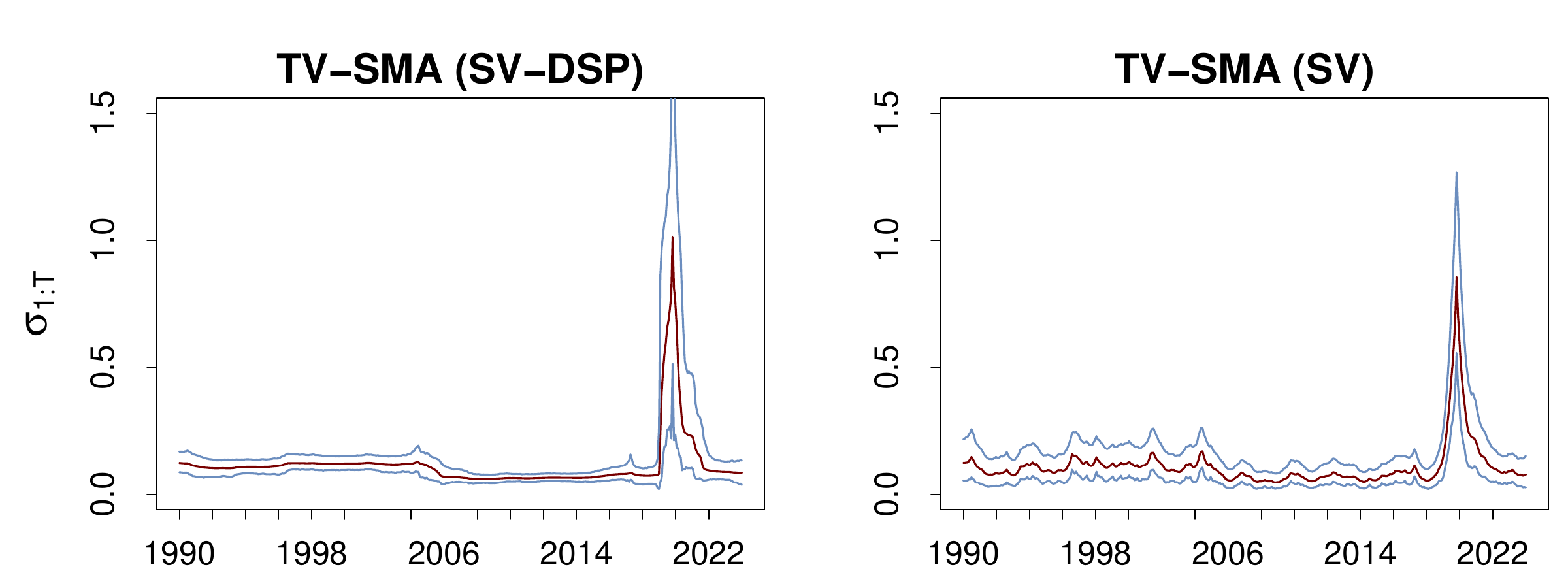}
   \caption{US air passenger data: The red and blue lines are posterior medians and $95\%$ HDIs over time for the posterior of the standard deviation of the observational noise process. }\label{fig:usair_noisevar_sma12}
\end{figure}

\begin{figure}
 \centering
 \includegraphics[width=1\textwidth]{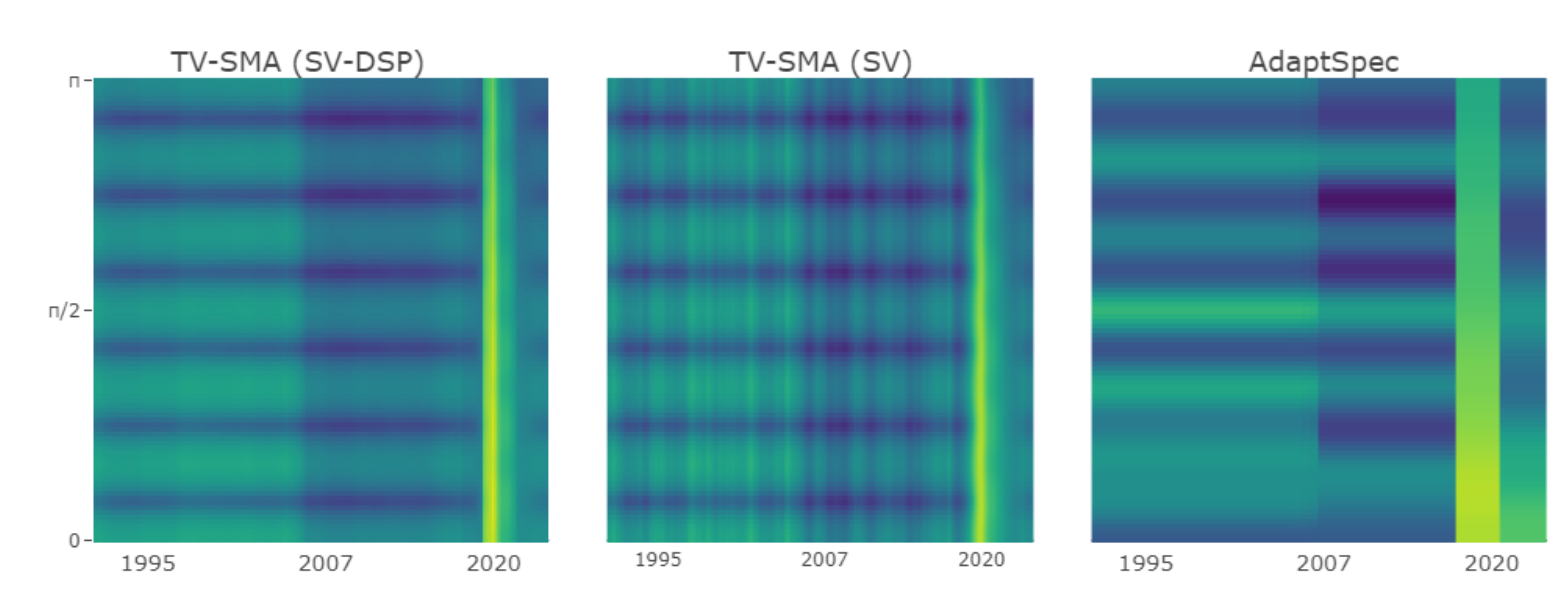}

    \caption{US air passenger data: posterior median spectograms from the TV-SMA$(1,2)_{12}$ (SV-DSP), TV-SMA$(1,2)_{12}$ (SV) and AdapSpec models}\label{fig:usair_SMA12_spectrograms}
\end{figure}

\begin{figure}[ht]
 \centering
 \includegraphics[width=0.8\textwidth]{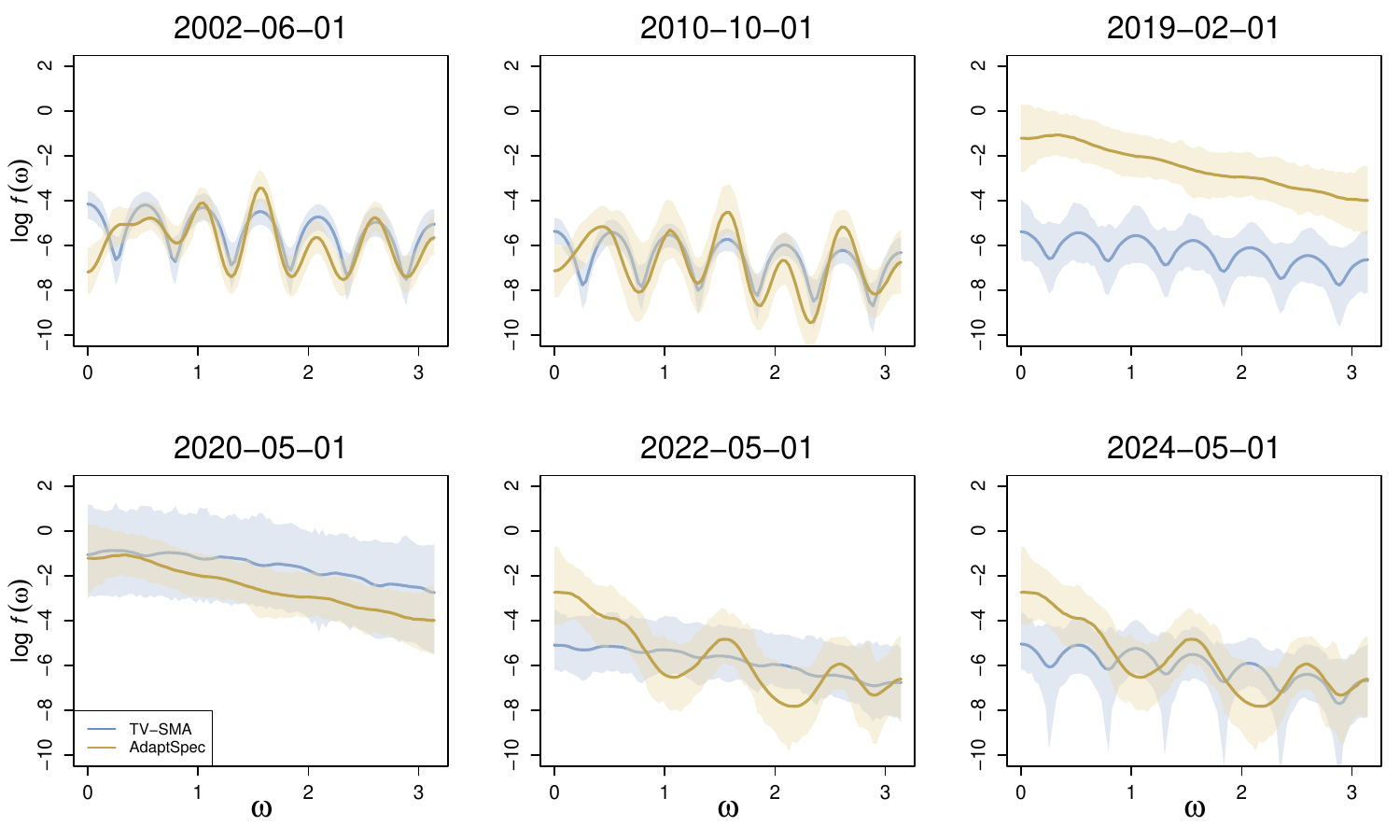}
    \caption{US air passenger data: Fitted spectral densities at six different time points from the TV-SMA(1,2)$(1,2)_{12}$ and AdaptSpec models. Solid lines are the posterior medians, and shaded bands are 95\% HDI. }\label{fig:usair_SAR12_spectral_snapshots}
\end{figure}
Both the TV-SMA$(1,2)_{12}$ (SV-DSP) and TV-SMA$(1,2)_{12}$ (SV) models yield nearly identical posterior distributions for the MA parameters in Figure \ref{fig:usair_sma12_coeff}. The  posteriors for the noise standard deviation from the two models, shown in  Figure \ref{fig:usair_noisevar_sma12}, exhibit a similar overall pattern, but the volatility from the SV-DSP model is much smoother than that of the SV model. Note also that the inferred noise process from the DSP prior exhibits a sharper spike during the COVID-19 period. In Figure \ref{fig:usair_noisevar_sma12}, the  vertical axis is truncated at 1.5 to facilitate comparison; the non-truncated distribution for the SV-DSP prior is provided in the Section \ref{app:usair_stability}.

Figure \ref{fig:usair_SMA12_spectrograms} presents heatmaps of the estimated time-varying median log spectral density for the TV-SMA$(1,2)_{12}$ (SV-DSP), TV-SMA$(1,2)_{12}$ (SV) and AdaptSpec. All three plots confirm the observations from Figure \ref{fig:usair_sma12_coeff}, namely that seasonal variation is substantially reduced during the Covid-19 pandemic period.

The spectrograms produced by the two TV-SMA$(1,2)_{12}$ models are similar, but the smoother posterior for the noise process in Figure \ref{fig:usair_noisevar_sma12} translates into a less wiggly  spectrogram from the TV-SMA$(1,2)_{12}$ (SV-DSP) model compared to the TV-SMA$(1,2)_{12}$ (SV) model. The slight level shift in noise process in Figure \ref{fig:usair_noisevar_sma12} around 2005 is clearly visible in the spectrograms, but is nowhere near the magnitude of the variance change during the pandemic. The spectrogram from AdaptSpec has the same overall pattern observed as the TV-SMA models, but the timing of the shifts are somewhat different and the big change actually happens before the start of the pandemic according to AdaptSpec. 

Figure \ref{fig:usair_SAR12_spectral_snapshots} provides snapshots of the spectral densities for selected time points for TV-SMA$(1,2)_{12}$ (SV-DSP) and AdaptSpec. Results for the TV-SMA$(1,2)_{12}$ (SV) model are omitted, as they are similar to those from the TV-SMA$(1,2)_{12}$ (SV-DSP) model. The panels in Figure \ref{fig:usair_SAR12_spectral_snapshots} are for six selected time points: three from the pre-pandemic period (top row) and three from the pandemic and post-pandemic period (bottom row), again showing substantially different inferred spectral densities from 2019 and onwards. 

\section{Conclusions}
Our article introduces a time-varying multi-seasonal ARMA model based on the exact likelihood, which extends the modeling framework in \cite{fagerberg2026time} for multi-seasonal AR models and the conditional likelihood. The model allows for multiple seasonal periods, a common feature in modern datasets, particularly those with high-frequency data such as hourly observations. The seasonalities are modeled with the parsimonious multiplicative structure in \citet{box2015time}. 

The time variation in the non-seasonal and seasonal ARMA coefficients are modeled by the DSP priors of \citet{kowal2019dynamic}.  At each time step, the AR and MA parameters are constrained to the stable AR and invertible MA regions,  preventing explosive behavior and ensuring identifiability, provided that the assumption of no root cancellation holds. The empirical results demonstrate that the model effectively captures extended periods of parameter stability, rapid changes and abrupt jumps.  

All model parameters are sampled from the joint posterior distribution using Gibbs sampling, building on the algorithm in \cite{fagerberg2026time}. To preserve the Markovian structure of the model and enable the use of the exact likelihood, the algorithm is extended by two additional updating steps for the pre-sample observations and the error path trajectory. The AR and MA parameters are sampled using the FFBSx sampler based on the EKF filter. \cite{fagerberg2026time}  demonstrate that the FFBSx sampler is fast, accurate and robust to the near-degeneracy in the state transition caused by the dynamic shrinkage process prior. Moreover, the gradient in the FFBSx algorithm can be  computed efficiently by automatic differentiation, making it straightforward to implement for any number of seasonal AR and MA polynomials. 

The proposed framework can be used with any of the recently developed global-local shrinkage priors, for example the prior processes in \citet{kalli2014time}, \citet{cadonna2020triple} and \citet{knaus2023dynamic}. 

The model and inference method are evaluated through two simulated examples: a pure ARMA model and a seasonal time-varying MA model. These are shown to perform well in comparison to the AdaptSpec method. The model is applied to a time series of the number of air passengers between Miami and New York, with a stochastic volatility model for the noise variance modeled by the DSP priors. The results reveal significant changes in seasonality and variance during the Covid-19 pandemic.

Future research should explore a careful modeling of a time-varying process mean in the model, including the potential interplay between time-varying ARMA parameters and the mean. An outlier component of the model can be introduced following \citet{wu2024trend}. It would interesting to assess the forecasting performance of the  model on multi-seasonal data, and to compare with alternatives. Finally, the framework can be extended to a seasonal vector ARMA model using the multivariate stability and invertibility restrictions in \citet{ansley1986note}, although care must be taken to handle the substantially larger number of parameters.

\if0\blind
{
\section*{Acknowledgments}
Mattias Villani was partially funded by the Swedish Research Council, grant 2025-01654. The computations were enabled by resources provided by the National Academic Infrastructure for Supercomputing in Sweden (NAISS), partially funded by the Swedish Research Council through grant agreement no. 2022-06725. The authors thank Héctor Rodríguez-Déniz for assisting in collecting the air passenger dataset.
}\fi

\bibliographystyle{apalike}
\bibliography{localarma_ref}

\newpage 


\title{ Supplement to \\ Time-Varying Multi-Seasonal ARMA Models}
\if0\blind{
\author{Ganna Fagerberg$^{a}$\thanks{Corresponding author: ganna.fagerberg@stat.su.se. $^a$Department of Statistics, Stockholm University. 
$^b$School of Business, University of New South Wales. $^c$Data Analytics Center for Resources and Environments (DARE).}, Mattias Villani$^{a}$ and Robert Kohn$^{b,c}$}
}\fi
\maketitle

\renewcommand{\theequation}{S\arabic{equation}}
\renewcommand{\thesection}{S\arabic{section}}
\renewcommand{\theproposition}{S\arabic{proposition}}
\renewcommand{\theassumption}{S\arabic{assumption}}

\renewcommand{\thethm}{S\arabic{thm}}

\renewcommand{\thelemma}{S\arabic{lemma}}
\renewcommand{\thealgocf}{S\arabic{algocf}}
\renewcommand{\thefigure}{S\arabic{figure}}
\renewcommand{\thetable}{S\arabic{table}}
\renewcommand{\thepage}{S\arabic{page}}
\renewcommand{\thetable}{S\arabic{table}}
\renewcommand{\thepage}{S\arabic{page}}
\setcounter{page}{1}
\setcounter{section}{0}
\setcounter{equation}{0}
\setcounter{algocf}{0}
\setcounter{lemma}{0}
\setcounter{assumption}{0}
\setcounter{table}{0}
\setcounter{figure}{0}
\setcounter{thm}{0}
\numberwithin{equation}{section}

\begin{abstract} 
    This Supplement contains additional results for the paper \emph{Time-Varying
Multi-Seasonal ARMA Models}.
\end{abstract}

\section{Additional results for Experiment 1}

\begin{figure} 
 \centering
 \includegraphics[width=0.8\textwidth]{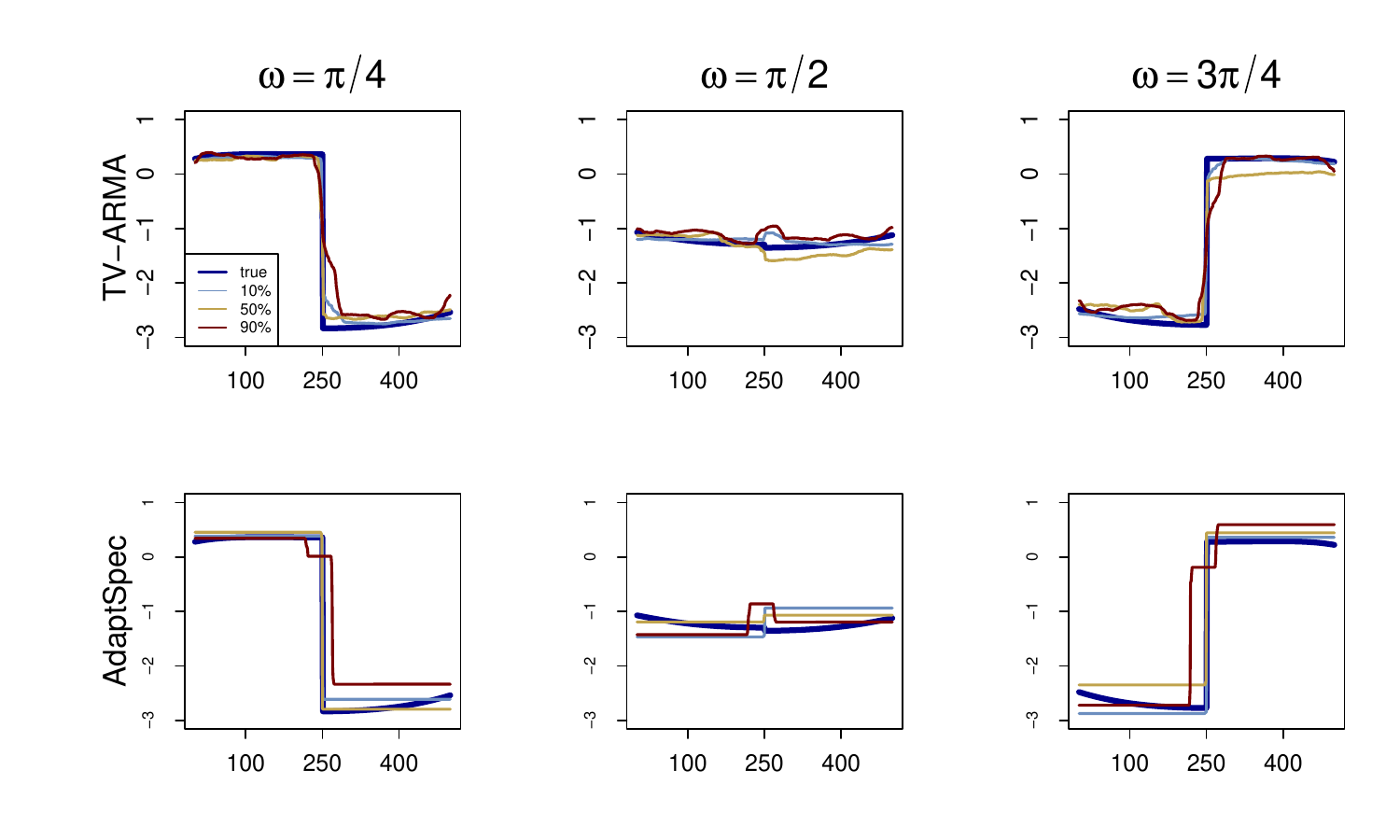}
   \caption{Experiment 1. Time evolution of the log spectral density at three different frequencies (columns) for several different model (rows). The dark blue line is the true spectral density and the three colored lines are the posterior median estimates from three different datasets chosen from the MSE percentiles.  }\label{fig:exp1spectra_time}
\end{figure}

\begin{figure} 
 \centering
 \includegraphics[width=0.8\textwidth]{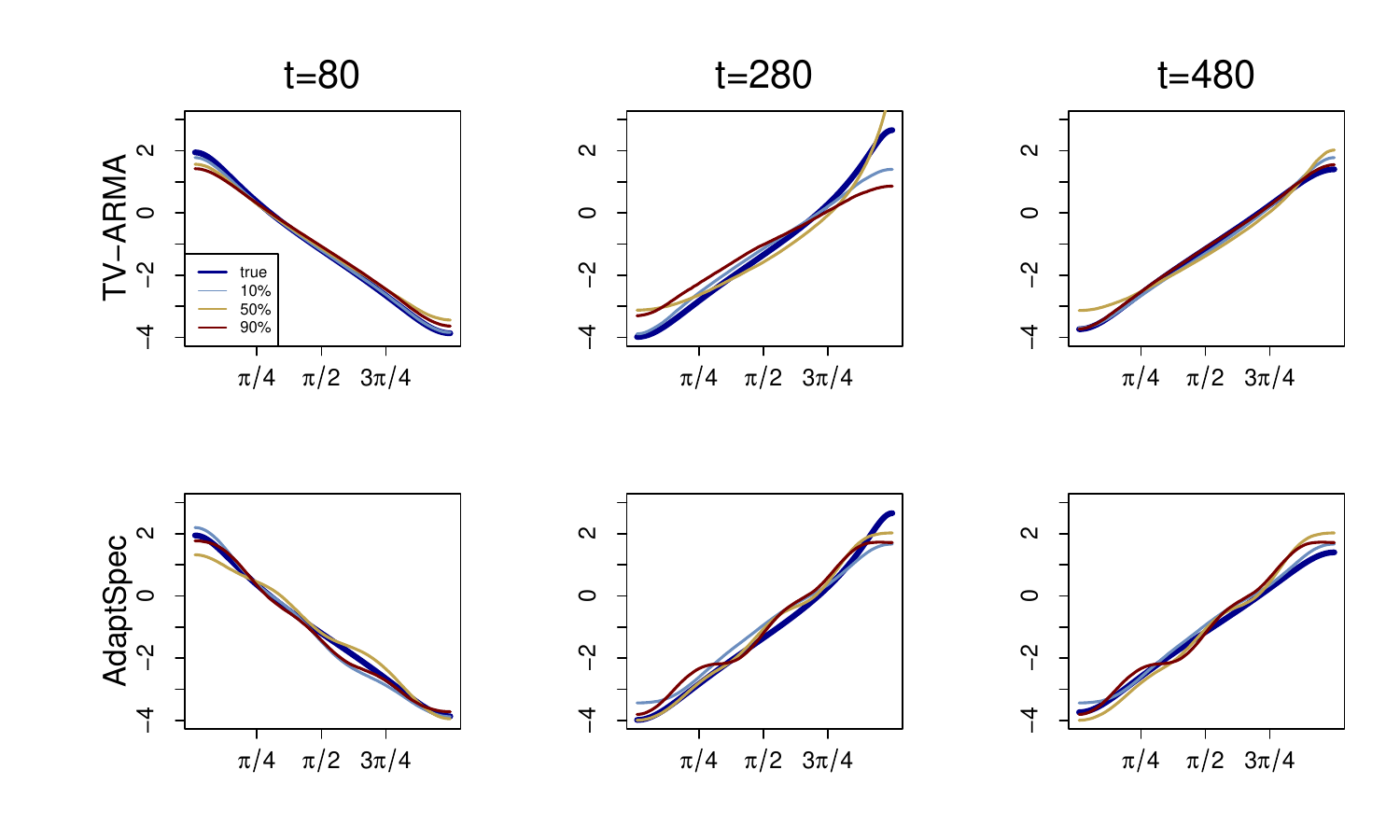}
   \caption{Experiment 1. Log spectral density at three different time points. The dark blue line is the true spectral density and the three colored lines are the posterior medians from three different datasets chosen from the MSE percentiles. }\label{fig:exp1spectra_freq}
\end{figure}

Figure  \ref{fig:exp1spectra_time} plots the estimated time-evolution at some selected frequencies for the models fitted to the datasets from Experiment 1. The graphs show the fit for three different datasets selected from the MSE percentiles for each method. Figure~\ref{fig:exp1spectra_freq} plots the fitted log spectral densities at three different time points, again showing the fit for three different datasets selected from the MSE percentiles for each method.

\section{US air passenger data - additional results }\label{app:usair_stability}

\begin{figure}
 \centering
 \includegraphics[width=1\textwidth]{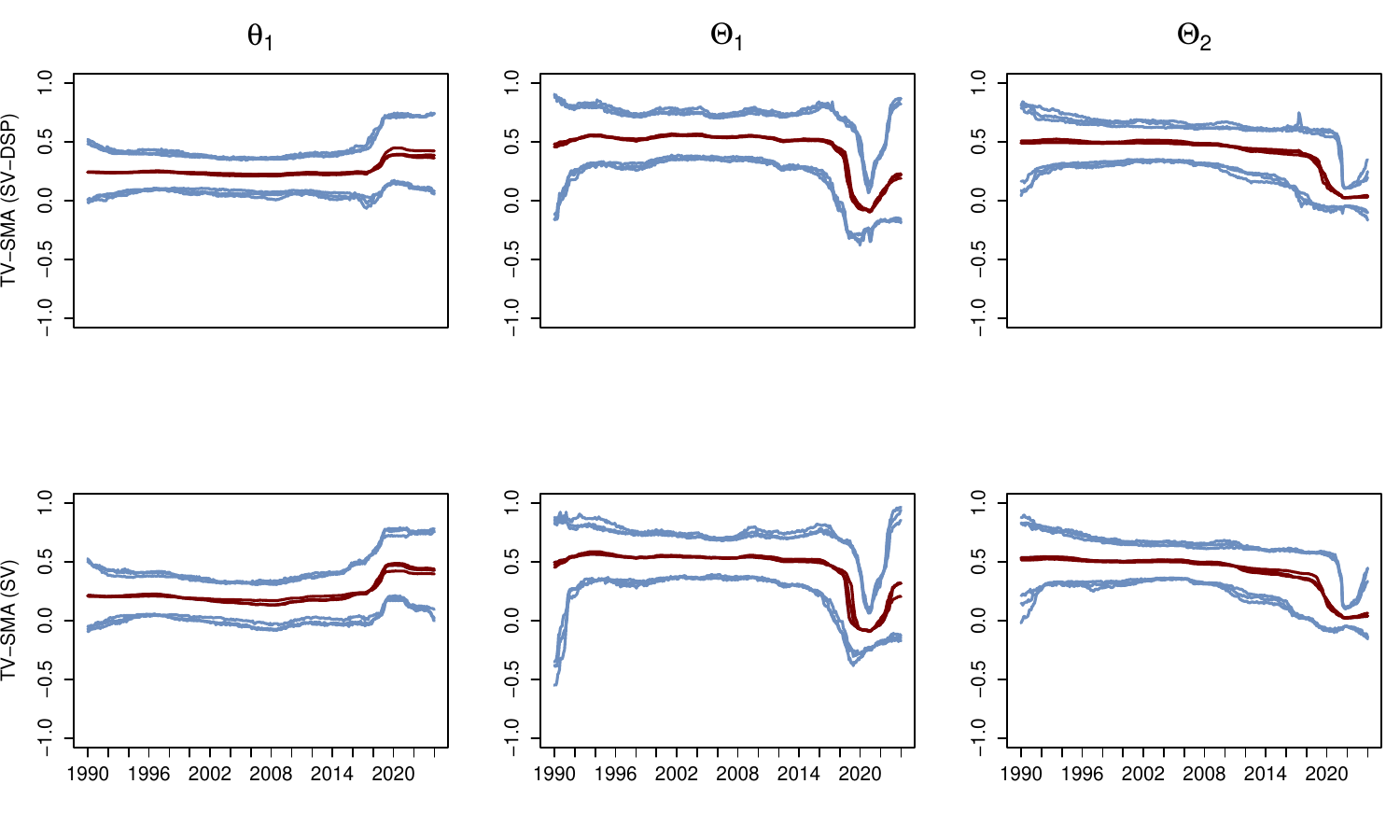}

   \caption{MCMC convergence of the  TV-SMA$(1,2)_{12}$ (SV-DSP) (top panel) and TV-SMA$(1,2)_{12}$ (SV) (bottom panel) models assessed by re-estimating the model using three different seeds and three different initial values for the global hyperparameter $\v\mu$. The red and blue lines are posterior medians and $95\%$ HDIs over time for each of the repeated runs. }\label{fig:usair_sma12_coef_stab}
\end{figure}

\begin{figure}
 \centering
 \includegraphics[width=0.8\textwidth]{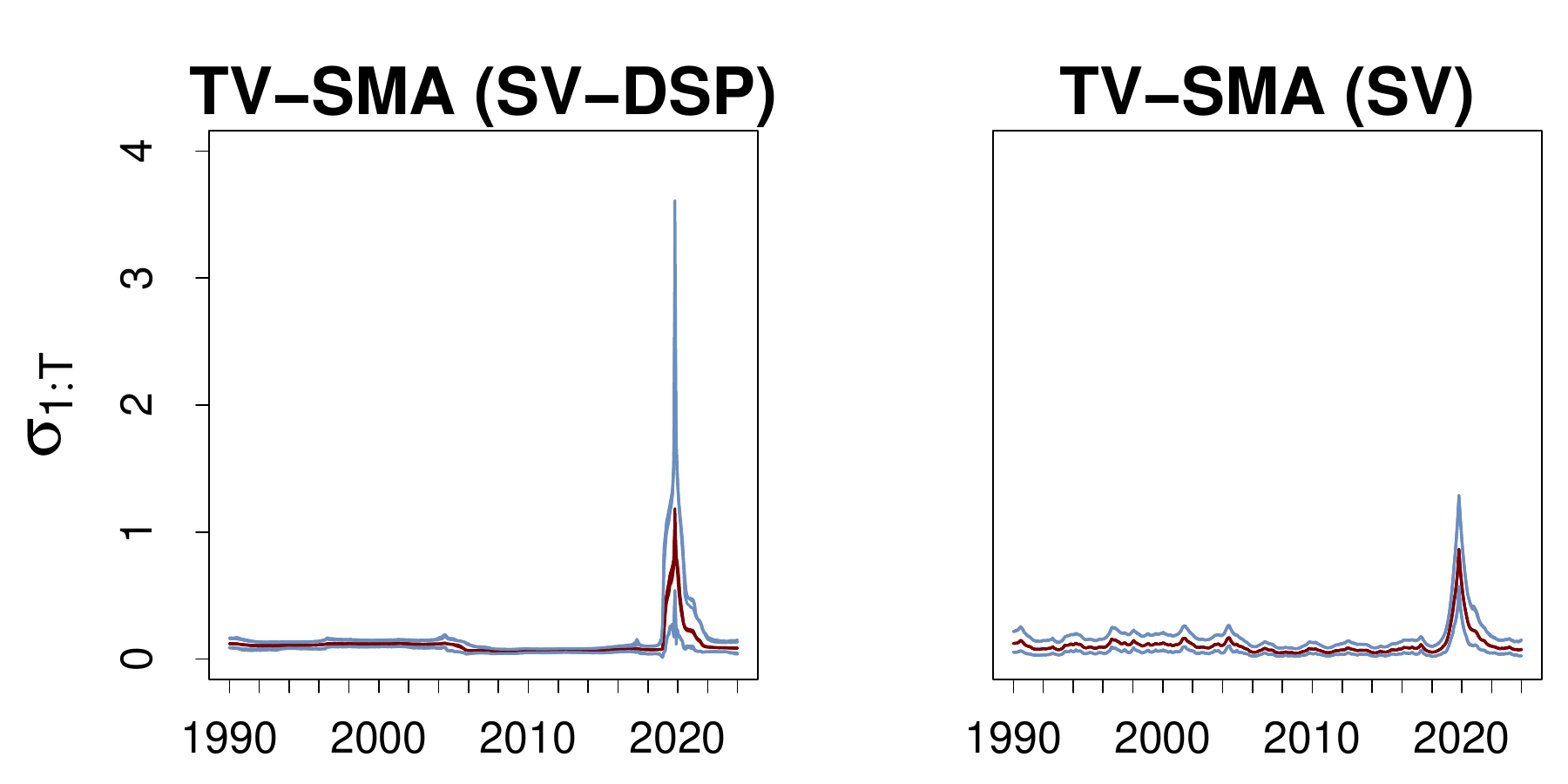}
   \caption{MCMC convergence of the posterior noise process for the TV-SMA$(1,2)_{12}$ (SV-DSP) (left plot) and TV-SMA$(1,2)_{12}$ (SV) (right plot)  models. The red and blue lines are posterior medians and $95\%$ HDIs over time for each of the five repeated runs}\label{fig:usair_sma_noise_stab}
\end{figure}

\begin{figure}
 \centering
 \includegraphics[width=0.8\textwidth]{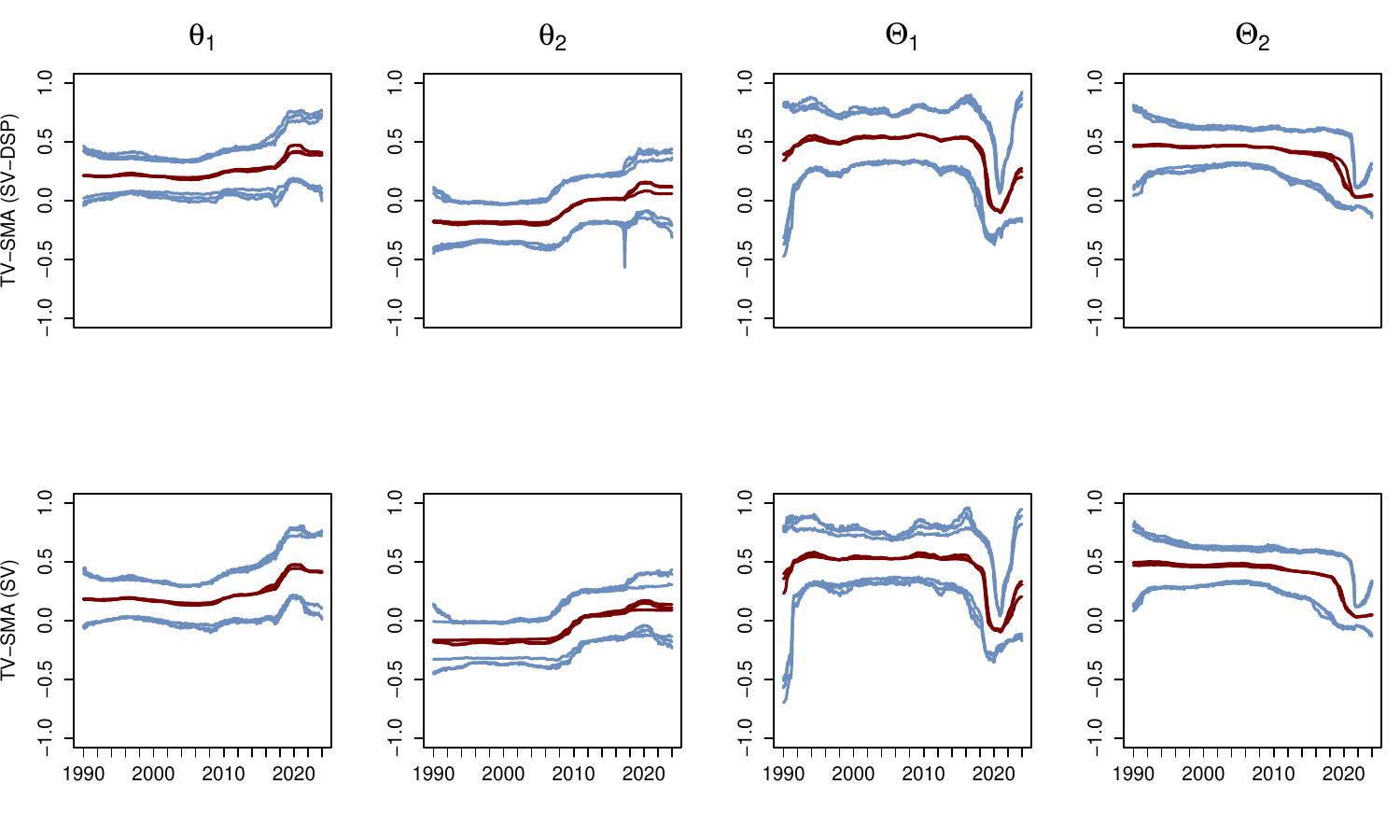}

   \caption{The MCMC convergence of the  TV-SMA($2,2)_{12}$ (SV-DSP) (top panel) and TV-SMA$(2,2)_{12}$ (SV) (bottom panel) models assessed by re-estimating the model using three different seeds and three different initial values for the global hyperparameter $\v\mu$. The red and blue lines are posterior medians and $95\%$ HDIs over time for each of the five repeated runs. }\label{fig:usair_sma22_coef_stab}
\end{figure}
\begin{figure} [h!]
 \centering
 \includegraphics[width=0.8\textwidth]{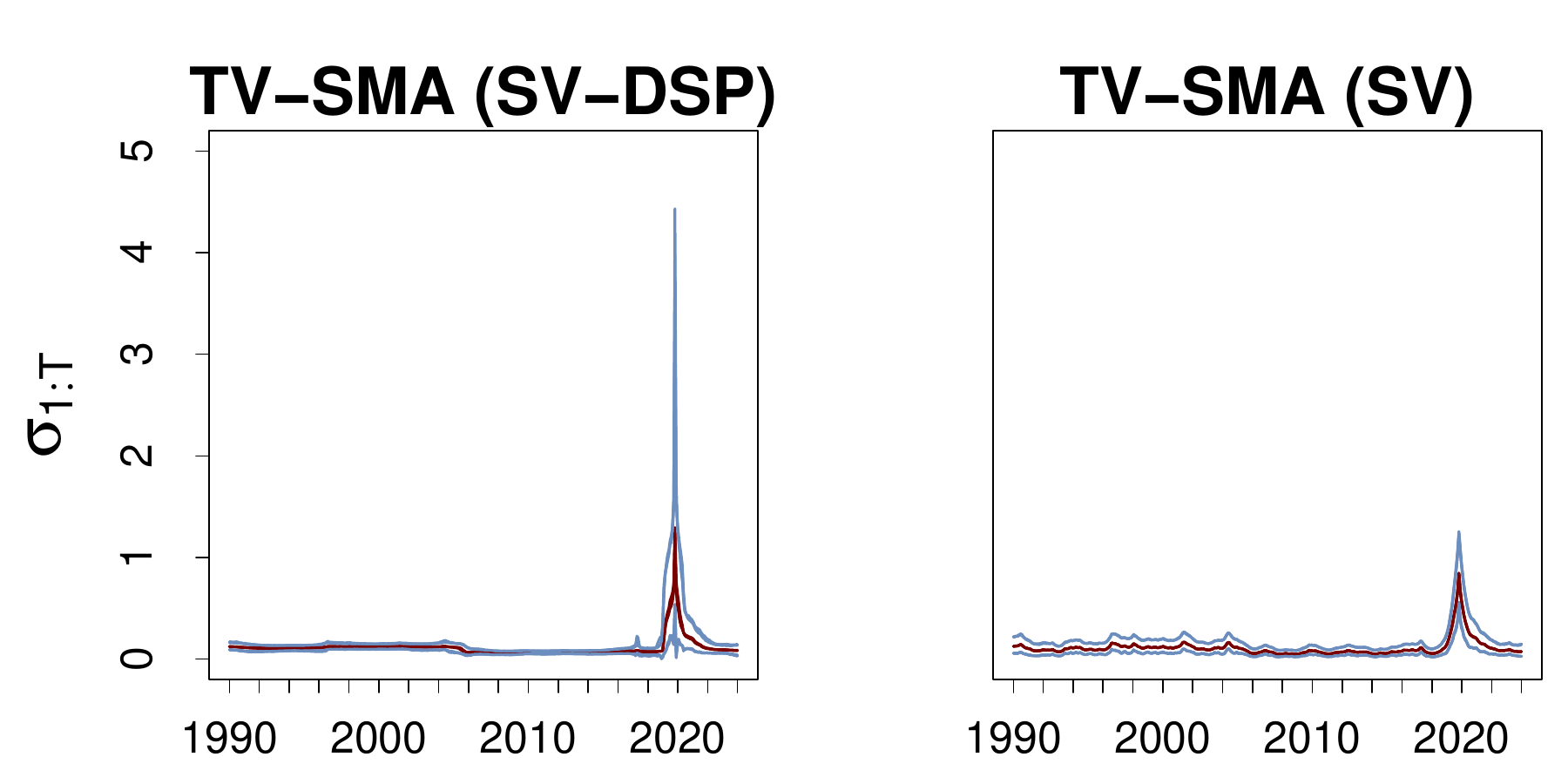}
   \caption{MCMC convergence of the posterior noise process for the TV-SMA$(2,2)_{12}$ (SV-DSP) and TV-SMA$(2,2)_{12}$ (SV)  models. The red and blue lines are posterior medians and $95\%$ HDIs over time for each of the five repeated runs}\label{fig:usair_sma22_noise_stab}
\end{figure}
\newpage

\end{document}